\documentclass[a4paper]{article}
\usepackage{geometry}
\usepackage{authblk}
\usepackage{lineno,hyperref}
\modulolinenumbers[5]

\usepackage{natbib}
\usepackage{epstopdf,graphicx,amssymb,amstext,amsmath,amsthm,color,multirow,mathtools,bbm,url}
\usepackage{subfigure}

\usepackage{threeparttable}
\usepackage{algorithm}
\usepackage{algpseudocode}

\newcommand\E{\mathbb{E}}
\newcommand\Prob{\mathbb{P}}
\newcommand\R{\mathbb{R}}

\newcommand\F{\mathcal{F}}
\newcommand\I{\mathcal{I}}

\newcommand\BS{\mathbf{S}}

\newcommand\M{\mathcal{M}}
\newtheorem{mydef}{Remark}

\newtheorem{mydef6}{Theorem}
\newtheorem{mydef7}{Lemma}

\usepackage{parskip}
\usepackage{mathtools,nccmath}%
\usepackage{ etoolbox, xparse} 
\DeclareMathOperator*{\argmin}{argmin}
\DeclarePairedDelimiterX{\abs}[1]\lvert\rvert{\ifblank{#1}{\,\cdot\,}{#1}}

\let\oldabs\abs
\def\abs{\futurelet\testchar\MaybeOptArgAbs}
\def\MaybeOptArgAbs{\ifx[\testchar\let\next\OptArgAbs
\else \let\next\NoOptArgAbs\fi \next}
\def\OptArgAbs[#1]#2{\oldabs[#1]{#2}}
\def\NoOptArgAbs#1{\ifblank{#1}{\oldabs{}}{\oldabs[\big]{#1}}}

\DeclarePairedDelimiterX{\set}[1]\{\}{\setargs{#1}}
\NewDocumentCommand{\setargs}{>{\SplitArgument{1}{;}}m}
{\setargsaux#1}
\NewDocumentCommand{\setargsaux}{mm}
{\IfNoValueTF{#2}{#1}{\nonscript\,#1\nonscript\;\delimsize\vert\nonscript\:\allowbreak #2\nonscript\,}}
\let\oldset\set
\def\set{\futurelet\testchar\MaybeOptArgSet}
\def\MaybeOptArgSet{\ifx[\testchar \let\next\OptArgSet
\else \let\next\NoOptArgSet \fi \next}
\def\OptArgSet[#1]#2{\oldset[#1]{#2}}
\def\NoOptArgSet#1{\OptArgSet[\big]{#1}}

\author[1, 2]{Vikranth Lokeshwar}
\author[1]{Vikram Bhardwaj}
\author[1]{Shashi Jain \thanks{shashijain@iisc.ac.in}}
\affil[1]{Department of Management Studies,\protect\\ Indian Institute of Science, Bangalore, India}
\affil[2]{EY GDS,\protect\\  Bangalore, India}
\title{Neural network for pricing and universal static hedging of contingent claims}
\begin{document}

\maketitle

\begin{abstract}
We present here a regress later based Monte Carlo approach that uses neural networks for pricing high-dimensional contingent claims.  The choice of specific architecture of the neural networks used in the proposed algorithm provides for interpretability of the model, a feature that is often desirable in the financial context. Specifically, the interpretation leads us to demonstrate that any contingent claim  ---possibly high dimensional and path-dependent--- under Markovian and no-arbitrage assumptions, can be semi-statically hedged using a portfolio of short maturity options. We show how the method can be used to obtain an upper and lower bound to the true price, where the lower bound is obtained by following a sub-optimal policy, while the upper bound by exploiting the dual formulation. Unlike other duality based upper bounds where one typically has to resort to nested simulation for constructing super-martingales, the martingales in the current approach come at no extra cost, without the need for any sub-simulations. We demonstrate through numerical examples the simplicity and efficiency of the method for both pricing and semi-static hedging of path-dependent options. 
\end{abstract}

\section{Introduction}

In this paper we consider the pricing and (semi) static hedging of derivative securities that include path-dependent options that are potentially based on several underlying assets. The efficient numerical evaluation of options written  on several underlying assets  can be challenging due to the so-called \emph{curse of dimensionality} which affects finite difference or tree-based methods. Given these restrictions Monte Carlo methods appear as attractive approach for the valuation of such derivatives.  

When it comes to pricing of path-dependent options, using the Monte Carlo approach, the main difficulty lies with the computation of the conditional expectation across different time intervals. The Least Squares Monte Carlo (LSMC) popularized by  \cite{carriere1996valuation}, \cite{tsitsiklis2001regression}, and   \cite{longstaff2001valuing} use the cross-sectional information from the simulated data to approximate the conditional expectation through a least squares regression. These methods generally fall in the category of the so-called \emph{regress now} approach, as in order to compute the conditional expectation at $t$ (\emph{now}), a least-squares regression of the value function at time $T>t$ is performed against basis functions of simulated data at time $t$ (\emph{now}). In contrast, with the \emph{regress later} approach, in order to compute the conditional expectation at $t,$ one first approximates the value function at time $T$ (\emph{later}) through a least squares regression using the basis functions of simulated data at time $T$ (\emph{later}). The basis functions in the \emph{regress later} approach is chosen such that their conditional expectations can be computed exactly which in turn results in the necessary conditional expectation at $t.$   \cite{glasserman2004simulation},  \cite{sgbm} are some of the examples of the \emph{regress later} approach. Analysis and discussion on regression later schemes can be found in  \cite{beutner2013fast} and \cite{balata2017}. More recently \cite{bender} propose the \emph{regression anytime},  which combines the \emph{regress later} and the \emph{regress now} approaches. In  \cite{rollingAdjoints} the \emph{regress later} approach is extended to compute path-wise forward Greeks.

There are a few advantages of using the \emph{regress later} over the \emph{regress now} approach. \cite{glasserman2004simulation} show that for a  single-period problem, the \emph{regress later} algorithm gives a higher coefficient of determination and a lower covariance matrix for the estimated coefficients.  \cite{pelsser} show that as the approximation error from the regression in the \emph{regress later} approach vanishes, the coefficients obtained are perfect regardless of the measure used for calibration. This property can be leveraged for problems where one has to work with mixed probability measures, examples of which include computing potential future exposures of Bermundan Swaptions in \cite{qian}, and computing the capital valuation adjustment in \cite{kva}, where one has to work simultaneously with both the risk-neutral and the real-world measures. A common problem faced by both, the \emph{regress now} and the \emph{regress later} approaches ---when a linear model is used for the regression--- is that the selection of the basis functions for the regression is arbitrary and varies between different payoffs. 

Static replication is an attractive alternative to the dynamic hedging of portfolios, as dynamic hedging often breaks down when there are sharp movements in markets or when the market faces liquidity issues.  Unfortunately, these are the precise moments where an effective hedge is highly desired. The principle of static replication is to construct a portfolio of instruments that mirrors the value function of a target security in every possible state of the world. Unlike dynamic hedging the portfolio weights of a static replicating portfolio do not change with changes in the market condition. By the no-arbitrage condition, if the value function of a target security is perfectly replicated at future time $T,$ the replicating portfolio should match the security’s value at an earlier time $t,$ ($t < T$) given no early exercise decision can be made between $t$ and $T.$  \cite{breeden} show that path-independent securities can be hedged using portfolio of standard options that mature along with the claim. For more general path dependent securities, in a single factor setting, good examples of static replication approach can be found in  \cite{derman}, and \cite{carr}.  \cite{pelsser2} utilizes the static portfolio replication to derive hedging strategies using swaptions for life insurance policies with guaranteed annuity options. For high-dimensional path-independent derivatives \cite{pellizzari} found that a hedging portfolio composed of statically held simple univariate options, optimally weighted minimizing the variance of the difference between the target claim and the approximate replicating portfolio resulted in a performance comparable to dynamic hedging, although only when there were significant transaction costs.

 There has been a recent boom in the financial applications of neural networks, mostly driven by the progress made in deep learning and the availability of specialized software and hardware. They have been used in \cite{liu}, \cite{liu2}, \cite{ackerer2019deep} and  \cite{bayer2019deep} for calibration of stochastic volatility and rough stochastic volatility models. Another strand of literature is on the applications of deep reinforcement learning, specifically related to data driven pricing and hedging of portfolios, which includes notably the work of  \cite{buehler} and  \cite{halperin}. Application of neural networks for model-based pricing of early exercise options includes the approach of policy iteration  in \cite{becker},  value function iteration in \cite{haugh}, \cite{kohler} and \cite{lapeyre2019neural}.  The work here falls under the category of value function iteration, although while others follow the \emph{regress now} framework for value function iteration, we here exploit the benefits of the \emph{regress later} approach.  
 
In this paper we develop a regress later based method for pricing high dimensional path dependent contingent claims. As the method uses a feed forward neural network for the regression at each time step, we call it \emph{Regress Later with Neural Network} (RLNN).  Specifically, RLNN uses a shallow network with two hidden layers ---with  rectified linear units (ReLU) activation functions used for the first hidden layer and linear activation function used for the second layer --- for the regression at each monitoring date. The choice of this basic architecture is deliberate, as it lends interpretability to the model, something valuable from a practitioner's point of view. Concisely, the first layer determines the payoff structure of a set of options, while the second layer determines the corresponding quantities of each of these options that needs to be held in a portfolio, so that the aggregated payoff of the resultant portfolio at any time $t \leq T$ replicates the value function of the target claim at time $t,$ where $T$ is the next monitoring date of the target claim.  
 
The semi-static replication using RLNN is not limited to single factor models but naturally applies to high dimensional contingent claims as well. RLNN therefore not only allows us to efficiently price a contingent claim, but also can be useful in determining an optimal static hedge, to avoid the practical difficulties and costs involved in dynamic hedging of the target claim. An advantage of RLNN over the traditional \emph{regress later} or \emph{regress now} approaches ---that rely on linear regression--- is that the method doesn't require a prior selection of basis functions depending upon the problem at hand.  A challenge that RLNN inherits from the \emph{regress later} schemes is the requirement for closed form or fast numerical approximations for the valuation of the short maturity options that are part of the static hedge portfolio. We demonstrate through examples that this issue can often be addressed in a relatively straightforward manner. 

There are three major contributions of this paper. First, we develop a method called Regress Later with Neural Networks (RLNN) for pricing high dimensional discretely monitored contingent claims. The second contribution is that we show any discretely monitored contingent claim ---written possibly on several underliers--- under the Markovian and no-arbitrage assumptions, can be semi-statically hedged using a basket of short maturity options. The  structure of the short maturity options, and their corresponding weights to be held in the hedge portfolio are the outcomes of the RLNN algorithm. Finally, we show that using RLNN one can obtain a tight lower and upper bound to the true price of the contingent claim at almost no extra costs. The lower bound is obtained by following the sub-optimal policy obtained by RLNN on a new set of paths. The upper bound is obtained using the dual formulation as prescribed by \cite{haugh} and \cite{rogers}. The major computational advantage with obtaining the duality based upper bound using RLNN comes from the fact that the required martingale can be obtained at no extra cost, which otherwise typically involves computationally heavy sub-simulations.  While \cite{glasserman2004simulation} point out the computational advantage of obtaining the duality based upper bounds for general \emph{regress later} algorithms, the quality of the upper bounds is left for future investigations. We through numerical experiments show that the quality of the upper and lower bounds obtained using RLNN are comparable or better than those reported in the literature.

The paper is organized as follows: Section \ref{section2} explains the notations used in the paper and the problem formulation. In Section \ref{section3} we describe the RLNN algorithm and the underlying neural network architecture. A universal semi-static hedging of contingent claims using the RLNN is discussed in  Section \ref{section4}. Section \ref{section4} also provides the proof of convergence of the RLNN method. Discussion on obtaining an upper and lower bound to the true price of the contingent claim using the RLNN method is presented in  Section \ref{section5}. In Section \ref{section6} through numerical examples we illustrate the accuracy of the method from both pricing and static hedging perspective.  Finally, we provide the conclusions in Section \ref{section7}. 

\section{Problem Formulation}\label{section2}
 This section defines the Bermudan option pricing problem and sets up the notations used in this paper. We assume a complete probability space $(\Omega,\mathcal{F},\mathbb{P})$ and finite time horizon $[0,T]$. $\Omega$ is the set of all possible realizations of the stochastic economy between 0 and $T$. 
The information structure in this economy is represented by an augmented filtration $\mathcal{F}_t:t\in[0,T],$  with
$\mathcal{F}_t$ the sigma field of distinguishable events at time $t$, and $\mathbb{P}$ is the risk-neutral probability measure on elements of $\mathcal{F}.$  
It is assumed that $\mathcal{F}_t$ is generated by $W_t$, a $d$-dimensional standard Brownian motion, and the state of economy is represented by an $\mathcal{F}_t$-adapted Markovian process, $\mathbf{S}_t = (S_t^1,\ldots,S_t^d) \in \mathbb{R}^d$, which has dependence on model parameters
 $\Theta = \{\theta_1,\ldots,\theta_{N_{\theta}}\}.$ We only consider discretely monitored contingent claims, i.e. the payoff of the claim depends only on the value of the price paths on the monitoring dates, $t\in[t_0 =0,\ldots,t_m,\ldots,t_M= T].$ Let $h_t:= h(\mathbf{S}_t)$ be an adapted process representing the intrinsic value of the option, i.e. the holder of the option receives $\max(h_t,0)$, if the option is 
exercised at time $t.$ With the risk-less savings account process, $B_t = \exp(\int_0^tr_s \, ds),$ where $r_t$ 
denotes the instantaneous risk-free rate of return, we define  $$D_{t_{m-1}} = \frac{B_{t_{m-1}}}{B_{t_{m}}}.$$ We consider the special case where $r_t$ is constant. The problem is then to compute
\begin{equation}\label{PathEstimator}
\frac{V_{t_0}(\mathbf{S}_{t_0})}{B_{t_0}} = \max_{\tau} \E\left[\frac{h(\mathbf{S}_{\tau})}{B_{\tau}} \right ],
\end{equation}
where $V_{t}(\cdot):t\in[0,T]$ is the option value function, and $\tau$ is a stopping time, taking values in the finite set $\{0,t_1,\ldots,T\}$. 

The dynamic programming formulation to solve this optimization problem is then as follows. The value of the option at the terminal time \emph{T} is equal to the product's pay-off,
\begin{equation}\label{E}
V_{T}(\mathbf{S}_T) = \max(h(\mathbf{S}_T),0).
\end{equation}
Recursively, moving backwards in time, the following iteration is then solved, given $V_{t_{m}}$ has already been determined, the continuation or hold value $Q_{t_{m-1}}$ is given by:
\begin{equation}\label{Q}
Q_{t_{m-1}}(\mathbf{S}_{t_{m-1}}) = B_{t_{m-1}}\E\left[\frac{V_{t_{m}}(\mathbf{S}_{t_{m}})}{B_{t_{m}}}\mid \mathbf{S}_{t_{m-1}}\right].
\end{equation}
The Bermudan option value at time $t_{m-1}$ and state $\mathbf{S}_{t_{m-1}}$ is then given by
\begin{equation} \label{V}
V_{t_{m-1}}(\mathbf{S}_{t_{m-1}}) = \max(h(\mathbf{S}_{t_{m-1}}),Q_{t_{m-1}}(\mathbf{S}_{t_{m-1}})).
\end{equation}

In a Monte Carlo simulation, the time evolution of the process $\mathbf{S}$ is approximated using some discretization scheme. We define a general Markov discretization scheme as

\begin{equation}\label{dis}
\mathbf{S}_{t_{m}} = F_{m-1}(\mathbf{S}_{t_{m-1}},\mathbf{Z}_{t_{m-1}},\Theta),
\end{equation}  

where $ \mathbf{Z}_{t_{m-1}}$ is a \emph{d-}dimensional standard normal random vector, and $F_{m-1}$ is a transformation from $\mathbb{R}^d$ to $\mathbb{R}^d.$ 

\section{Regress later with Neural networks (RLNN)}\label{section3}

The algorithm begins by generating $N$ independent copies of sample paths, $\{\mathbf{S}_{t_0},\ldots,\mathbf{S}_{t_M}\},$ of the underlying process that are obtained using the recursion,
$
\mathbf{S}_{t_{m}}(n) = F_{m-1}(\mathbf{S}_{t_{m-1}}(n), \mathbf{Z}_{t_{m-1}}(n), \Theta), 
$
where $n = 1,\ldots, N$ is the index of the path. 

The method then computes the value of the contingent claim at terminal time as
$V_{t_M}(\mathbf{S}_{t_M}) = \max(h(\mathbf{S}_{t_M}),0)$.

The following steps are then employed for each monitoring date, $t_m,\,m\leq M,$ recursively moving backwards in time, starting from $t_{M}$:

Assume that  $\tilde{V}_{t_m}(\mathbf{S}_{t_m}(n)),\ n = 1,\ldots, N,$ our estimates for  $V_{t_m}(\mathbf{S}_{t_m}(n)),$ are known. 

\begin{itemize}

\item{\bf \emph{Regress-later} }\label{SecRegLater}

In this step a parametrized value function $\tilde{G}:\R^d \times \R^{N_p} \mapsto\R,$ which assigns values $\tilde{G}(\mathbf{S}_{t_m},\beta_{t_m})$ to states $\mathbf{S}_{t_m},$ is computed. Here $\beta_{t_{m}}\in \R^{N_p}$ is a vector of free parameters.  The objective is to choose for each $t_m,$  a parameter vector $\beta_{t_{m}},$ so that
\begin{equation*}
\tilde{G}(\mathbf{S}_{t_m},\beta_{t_{m}})= \tilde{V}_{t_m}(\mathbf{S}_{t_m}).
\end{equation*}

By the universal approximation theorem, for appropriate choice of activation functions, a neural network with two layers and sufficiently large number of neurons can approximate any continuous function over a compact set arbitrarily well (see  \cite{hornick}). We choose as our approximation architecture at $t_m$ a feed-forward network $\tilde{G}^{\beta} : \R^d \rightarrow\R$ of the form

\begin{equation*}
\tilde{G}^{\beta_{t_m}} := \psi \circ A_2 \circ \varphi \circ A_1
\end{equation*}
where $ A_1:\R^d \rightarrow \R^p$ and $ A_2:\R^p\rightarrow\R$ are affine functions of the form,

\begin{equation*}
A_1(\mathbf{x}) = \mathbf{W}_1\mathbf{x} + \mathbf{b}_1 \ \ \textrm{for} \ \mathbf{x} \in \R^d,\ \mathbf{W}_1 \in \R^{p \times d}, \mathbf{b}_1 \in \R^p,
\end{equation*}
and
\begin{equation*}
A_2(\mathbf{x}) = \mathbf{W}_2\mathbf{x} + b_2 \ \ \textrm{for} \ \mathbf{x} \in \R^p,\ \mathbf{W}_2 \in \R^{1 \times p}, b_2 \in \R.
\end{equation*}

$\varphi : \R^j \rightarrow \R^j, j \in \mathbb{N}$ is the component-wise ReLU activation function given by:
\begin{equation*}
\varphi(x_1,\ldots,x_j):=\left(\max(x_1,0),\ldots,\max(x_j,0)\right),
\end{equation*}

while $\psi:\R^j \rightarrow \R^j, j \in \mathbb{N}$ is the component-wise linear activation function given by:
\begin{equation*}
\psi(x_1,\ldots,x_j):=\left(x_1,\ldots,x_j\right).
\end{equation*}

The dimension of the parameter space is therefore,
\begin{equation*}
N_p = 1+p+p+p \times d.
\end{equation*}

Specifically, $\beta_{t_{m}}\in \R^{N_p}$ is chosen to minimize the following mean squared error.

\begin{equation}
\beta_{t_m} = \argmin_{\beta_{t_m}}\left(\frac{1}{N} \sum_{n = 1}^{N}\left(\tilde{V}_{t_m}(\mathbf{S}_{t_m}(n))-\tilde{G}^{\beta_{t_m}}\left(\mathbf{S}_{t_m}(n)\right)\right)^2\right),
\end{equation}

where $\mathbf{S}_{t_m}(n),\ n = 1,\ldots,N$ now constitute the set of training points.

Figure \ref{fig1} presents as schematic diagram of the neural network at $t_m.$

\begin{figure}
\center
\includegraphics[width = 0.8\textwidth]{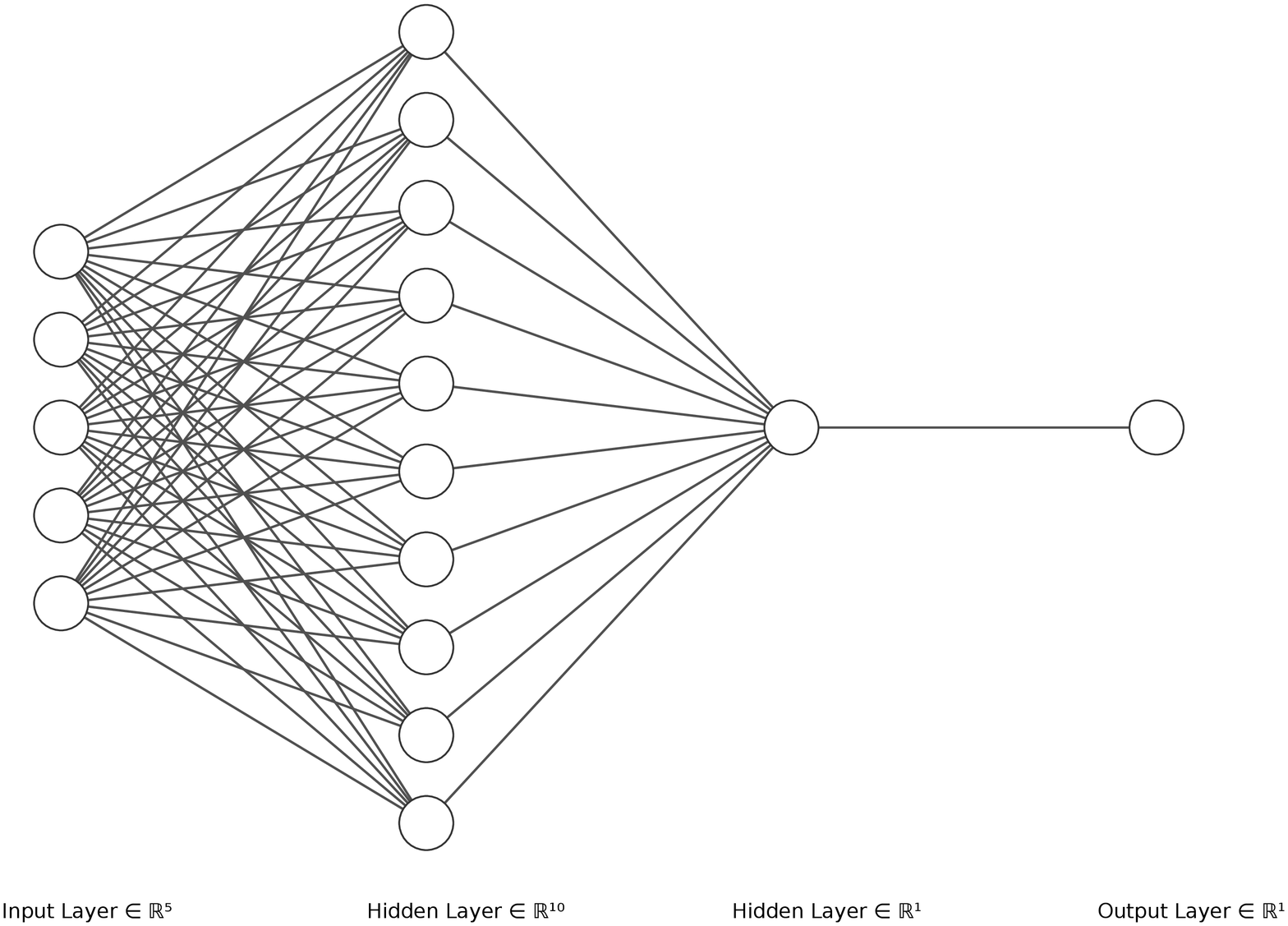} 
\caption{The neural network architecture chosen for each monitoring date, where the input dimension $d$ = 5, the number of hidden neurons $p$ in the first hidden layer is 16 and the second hidden layer is equal to 1. } \label{fig1}
\end{figure}

\item{ \bf Computing the continuation and option values at $t_{m-1}$}

The continuation values for the sample points $\BS_{t_{m-1}}(n), \ n = 1,\ldots,N $ are then approximated by, 
\begin{eqnarray}\nonumber
\widehat{Q}_{t_{m-1}}\left(\BS_{t_{m-1}}(n)\right) &=&  B_{t_{m-1}}\E\left[\frac{\tilde{V}_{t_{m}}(\mathbf{S}_{t_{m}})}{B_{t_{m}}}\mid \mathbf{S}_{t_{m-1}}(n)\right]\\ \nonumber
&\approx & B_{t_{m-1}}\E\left[\frac{\tilde{G}^{\beta_{t_m}}\left(\mathbf{S}_{t_m}\right)}{B_{t_{m}}}\mid \mathbf{S}_{t_{m-1}}(n)\right]\\\label{optPortEqn}
\end{eqnarray}

Like the other \emph{regress-later} based schemes the method requires, either a closed form solution, or a fast-numerical approximation for \\ $\E\left[\frac{\tilde{G}^{\beta_{t_m}}\left(\mathbf{S}_{t_m}\right)}{B_{t_{m}}}\mid \mathbf{S}_{t_{m-1}}(n)\right].$ Later in Section \ref{section4} we provide for our specific choice of network architecture a static hedging interpretation and show that Equation \ref{optPortEqn} boils down to evaluating the price of short maturity options that are written at $t_{m-1}$ and expire at $t_m.$  In Section \ref{section6} we show with the help of practical examples how the above expectation can usually be readily evaluated. 

Once the approximate continuation value $\widehat{Q}_{t_{m-1}}\left(\BS_{t_{m-1}}(n)\right)$ for samples $\BS_{t_{m-1}}(n), \ n = 1,\ldots,N$ have been computed the corresponding approximation for the option values are given by

\begin{equation}\label{vIteration}
\tilde{V}_{t_{m-1}}(\mathbf{S}_{t_{m-1}}(n)) = \max\left(h\left(\BS_{t_{m-1}}(n)\right),\widehat{Q}_{t_{m-1}}\left(\BS_{t_{m-1}}(n)\right)\right).
\end{equation}
\end{itemize}

The method discussed above is summarized as Algorithm \ref{alg1} below. 

\begin{algorithm}
\caption{Regress Later with Neural Networks (RLNN)}\label{alg1}
\begin{algorithmic}[1]
\State Generate $\BS_{t_m}\left(n\right)$  for paths $n=1,\ldots,N, \, m = 0,\ldots,M$ 
\State $\tilde{V}_{t_M} \leftarrow h \left(\BS_{t_M}\right)$  evaluate final time option value for each path.
\State Initialize $\beta_{t_M}$ from uniform distribution. 
\For{$m=M-1\ldots,1$}
       \State  $\beta_{t_m} \leftarrow \argmin_{\beta_{t_m}}\left(\frac{1}{N} \sum_{n = 1}^{N}\left(\tilde{V}_{t_m}(\mathbf{S}_{t_m}(n))-\tilde{G}^{\beta_{t_m}}\left(\mathbf{S}_{t_m}(n)\right)\right)^2\right) $ fitting the network
\For{$n=1,\ldots,N$}       
       \State $\widehat{Q}_{t_{m-1}}\left(\BS_{t_{m-1}}(n)\right) \leftarrow B_{t_{m-1}}\E\left[\frac{\tilde{G}^{\beta_{t_m}}\left(\mathbf{S}_{t_m}\right)}{B_{t_{m}}}\mid \mathbf{S}_{t_{m-1}}(n)\right]$ the estimated continuation value
       \If{$h(\BS_{t_{m-1}}(n)) > \widehat{Q}_{t_{m-1}}\left(\BS_{t_{m-1}}(n)\right)$}
    			\State$\tilde{V}_{t_{m-1}}(\BS_{t_{m-1}}(n)) \leftarrow h(\BS_{t_{m-1}}(n))$
  	  \Else
    			\State $\tilde{V}_{t_{m-1}}(\BS_{t_{m-1}}(n)) \leftarrow \widehat{Q}_{t_{m-1}}\left(\BS_{t_{m-1}}(n)\right)$
      \EndIf
      \EndFor
      \State  $\beta_{t_{m-1}} \leftarrow \beta_{t_{m}}$ initialize the parameters for network to be trained at $t_{m-1}$
    \EndFor
 
\end{algorithmic}
\end{algorithm}

\section{Static hedging with RLNN}\label{section4}

Before we elaborate on the static hedging interpretation of RLNN we briefly discuss a closely related static hedging approach proposed by \cite{carr}. They show that the time-$t$ value of a European option maturing at a fixed time $T \geq t$ relates to the time-$t$ value of a continuum of shorter maturity European call options.

Let $V_t^T(\BS_t,K)$ denote the time $t$ value of a vanilla European call option with strike $K$ that matures at $T.$ Consider a class of models in which the risk-neutral evolution of the asset price follows a single factor Markov process, together with the assumption that the markets are frictionless and there is no arbitrage (there exists a risk-neutral probability measure $\Prob$).  It is also assumed that the vector of model parameters $\Theta$ are fixed.\\

\begin{mydef6} (\cite{carr})
Under the Markovian and no-arbitrage assumptions, the time-$t$ value of a European option maturing at a fixed time $T \geq t$ relates to the time-$t$ value of a continuum of European call options at a shorter maturity $u \in \left[t,T\right],$ by
\begin{equation}
V_t^T(S_t,K) = \int_0^{\infty} w(S_u) V_t^u(S_t,S_u) \,dS_u,\ u \in [t,T],
\end{equation}
for any $S_t >0$ and  $t \leq u.$ The weights $w(S_u)$ do not vary with $S_t$ or $t,$ and are given by
\begin{equation}\label{wts}
w(S_u) = \frac{\partial^2}{\partial S_u^2} V_u^T(S_u,K)
\end{equation}
\end{mydef6}

The outline of the proof is as follows:

  Under the Markovian assumption, and using law of iterated conditioning

\begin{eqnarray}\nonumber
V_t^T(\BS_t,K) &=&e^{-r(u-t)} \E\left[e^{-r(T-u)}\E\left[\left(\BS_T-K\right)^+\mid\F_u\right]\mid\F_t\right], \ u \in \left[t,T\right]\\ 
&=& e^{-r(u-t)} \E\left[V_u^T(\BS_u,K)\mid\F_t\right], \label{markovV}
\end{eqnarray}

where for simplicity a constant risk-free rate $r$ has been assumed.

For a single factor Markovian process, using the results of  \cite{breeden}, who show the risk neutral density can be equated to the second strike derivative of the call option using the following relation:

\begin{equation}\label{densityEqn}
\Prob\left(S_T \mid S_t\right)= e^{r(T-t)} \frac{\partial^2 }{\partial S_T^2}V_t^T\left(S_t, S_T\right)
\end{equation}

Therefore,   Equation \ref{markovV} can be re written as,

\begin{eqnarray}\nonumber
V_t^T(S_t,K) &=& e^{-r(u-t)} \E\left[V_u^T(S_u,K)\mid\F_t\right], \ u \in \left[t,T\right]\\ \nonumber
&=& e^{-r(u-t)} \int_0^{\infty} V_u^T(S_u,K) \Prob(S_u|S_t)\, dS_u \\ 
&=&  \int_0^{\infty} V_u^T(S_u,K)  \frac{\partial^2 }{\partial S_u^2}V_t^u\left(S_t, S_u\right)\, dS_u \label{int1} \\
&=& \int_0^{\infty} V_t^u(S_t,S_u) \frac{\partial^2 }{\partial S_u^2} V_u^T(S_u,K)\label{int2},
\end{eqnarray}

with the density function in Equation \ref{int1} substituted by the relation provided in Equation \ref{densityEqn}. Equation \ref{int2} is the outcome of twice integration by parts with appropriate boundary conditions. 

The hedging interpretation of Theorem 1 is, in order to hedge a target European call option that matures at $T,$ at $t < T$ one enters into a continuum of positions in options that mature at $u <T,$ with the weights of the positions given by the relation in Equation \ref{wts}. As the weights do not depend upon the asset price $S_t$ or $t$ the hedge is static over the period $[t, u].$ At $u$ the proceeds from the payoff of the maturing options is just enough to either close the position of the target option or set up another hedge. It should be noted that when the positions are rolled over at $u,$ there is a risk that the Markov condition might not hold, the $\Theta$ parameters would change, and therefore there lies a model risk. 

As holding a continuum of option positions is not feasible, ideally one would hold a finite set of options. \cite{carr} propose the use of Gaussian quadrature to optimally span positions in finite number of options. The method can also be extended to path-dependent options where the value of the contingent claim depends upon a finite number of points on the price path of single underlying asset. The discrete time points at which the price of underlying assets determines the value of the contingent claim are labelled as the monitoring times, a notion that we also follow in this paper. 

While Theorem 1 states that under assumptions the value function of a longer maturity contingent claim can be replicated by setting up a static portfolio comprising of short maturity options there are practical challenges in extending the same for contingent claims whose value depends on several underlying assets. We here show that using the RLNN one can effectively determine the semi-static hedge for high-dimensional contingent claims, including discretely monitored path dependent claims. 

We begin by providing interpretation of t neural network used in the RLNN followed by demonstrating how it can be used for semi-static hedging on contingent claims under model assumptions.

\subsection{Interpretation of the first hidden layer}

The outcome of the first hidden layer with a choice of $p$ neurons can be represented as 

\begin{equation*}
\mathbf{o} = \varphi\left(\mathbf{W}_1\mathbf{x} + \mathbf{b}_1 \right), \  \mathbf{x} \in \R^d,\ \mathbf{W}_1 \in \R^{p \times d}, \mathbf{b}_1 \in \R^p,
\end{equation*}

where $\varphi : \R^p \rightarrow \R^p, p \in \mathbb{N}$ is the component-wise ReLU activation function.

Each of the $p$ elements of $\mathbf{o}:=\left\{o_1,\ldots,o_p \right\}$ therefore has the form,

\begin{equation*}
o_i = \max\left(\sum_{j = 1}^{d} w_{ij}x_j + b_{i1},0\right),
\end{equation*}

where $w_{ij}, \ j = 1,\ldots,d$  is the $i$th row of $\mathbf{W}_1,$
\begin{equation*}
\mathbf{W}_1 := \begin{bmatrix}
    w_{11} & w_{12} &  \dots  & w_{1d} \\
    w_{21} & w_{22}  & \dots  & w_{2d} \\
    \vdots & \vdots & \ddots & \vdots \\
    w_{p1} & w_{p2}  & \dots  & w_{pd}
\end{bmatrix} := \begin{bmatrix} \mathbf{w}^{\top}_1 \\
                                                         \mathbf{w}^{\top}_2 \\
                                                         \vdots\\
                                                         \mathbf{w}^{\top}_p
                                                         \end{bmatrix}
\end{equation*}

It is easy to see that the $i$th neuron, where $ i = 1,\ldots,p,$ has the form of the payoff of an arithmetic basket option with weights $w_{ij}, \ j = 1,\ldots,d$ and strike  $b_{i1},$ written on the underlying $\mathbf{x}:=\{x_1,\ldots,x_d\}.$ Therefore, the outcome of the first hidden layer provides the structure, i.e. the composition of the weights and the strike, of a set of $p$ arithmetic basket options. 

\subsection{Interpretation of the second hidden layer}

The second hidden layer performs the following operation:

\begin{equation}
y = \left(\sum_{i = 1}^p \omega_{i} o_i \right) +b_2, \ \text{where,} \ \mathbf{W}_2 := \{\omega_1,\ldots,\omega_p\},
\end{equation}

which can be seen as determining the weights of the $p$ basket options you need to hold in your portfolio. The amount you need to invest  in the risk-free asset is given by the bias of the second hidden layer, i.e. $b_2.$ 

Under the assumption that the asset price follows a Markovian process, the market is frictionless and there is no arbitrage, we next show that the portfolio of options constructed as above replicates the contingent claim, between monitoring dates. We additionally assume that  $\BS_{t_m} \in \I_d,$ for $ m = 0,\ldots, M,$ almost surely, where $\I_d$ is some d-dimensional hypercube. As a consequence, we assume,  $V_{t_m} \in C(\I_d),$ where $C(\I_d)$ denotes the space of real-valued continuous functions on $\I_d.$ Note that for ease of notation, we here after drop out the discounting terms.\\

\begin{mydef7}\label{lemma1}
Given $\epsilon >0$ there exists $p>0$ such that 
\begin{equation}
\tilde{G}^{\beta_{t_m}}(\mathbf{x}) = \sum_{i = 1}^p \omega_{i} \varphi \left(\mathbf{w}^{\top}_{i}\mathbf{x} + b_{i} \right)+b_2,
\end{equation}

 satisfies
 
 \begin{equation}
 \sup_{\mathbf{x} \in \I_d} \left| V_{t_m} (\mathbf{x}) - \tilde{G}^{\beta_{t_m}}(\mathbf{x})\right| < \epsilon
 \end{equation}

\end{mydef7}

\emph{Proof:} Lemma \ref{lemma1} is the direct consequence of the extension of universal approximation theorem of \cite{hornick} by \cite{leshno}, who show that an activation function will lead to a network with universal approximation capacity, if and only if, the function (in our case $\varphi$) is not a polynomial almost everywhere. \\

\begin{mydef7} \label{lemma2}
Under the Markovian and no arbitrage assumption, and for sufficiently large $p,$ the value of the contingent claim at $t \in (t_{m-1},t_m]$ holds the following relation
\begin{equation}
\sup_{\BS_t \in \I_d}\left|V_t(\BS_t) - \E\left[\tilde{G}^{\beta_{t_m}}(\BS_{t_m} )\mid \BS_t \right]\right| < \epsilon
\end{equation}
where, $D_t$ is the discounting factor, and $\epsilon>0$ is given. 
\end{mydef7}

\emph{Proof:} Lemma \ref{lemma2} is the result of the Markovian and the no-arbitrage assumption together with the use of Lemma \ref{lemma1}.

\begin{eqnarray}\nonumber
\left|V_t(\BS_t) -\E\left[\tilde{G}^{\beta_{t_m}}(\BS_{t_m}) \mid \BS_t \right]\right| &=& \left|\E\left[V_{t_m}\left(\BS_{t_m}\right)\mid\BS_t\right] -\E\left[\tilde{G}^{\beta_{t_m}}(\BS_{t_m}) \mid \BS_t \right]\right|,\\ \nonumber
&=& \left|\E\left[V_{t_m}\left(\BS_{t_m}\right)-\tilde{G}^{\beta_{t_m}}(\BS_{t_m}) \mid \BS_t \right]\right|,\\ \nonumber
&\leq & \E\left[\abs{V_{t_m}\left(\BS_{t_m}\right)-\tilde{G}^{\beta_{t_m}}(\BS_{t_m})} \mid \BS_t \right],\\ \nonumber
& < & \E\left[\epsilon \mid \BS_t \right] = \epsilon,
\end{eqnarray}

where the first line is the consequence of the Markovian assumption, the second line is from the no-arbitrage assumption and use of the linearity of the conditional expectations. The third line uses the Jensen's inequality, while the last line is the outcome of Lemma \ref{lemma1}.

The evaluation of $\E\left[\tilde{G}^{\beta_{t_m}}(\BS_{t_m}) \mid \BS_t \right],$ can be expanded as follows:

\begin{eqnarray}\nonumber
\E\left[\tilde{G}^{\beta_{t_m}}(\BS_{t_m}) \mid \BS_t \right] &=& \E\left[\sum_{i = 1}^p \omega_{i} \varphi \left(\mathbf{w}^{\top}_{i}\BS_{t_m} + b_{i} \right)+b_2\mid \BS_t\right]\\
&=& \sum_{i = 1}^p \omega_{i} \E\left[\varphi \left(\mathbf{w}^{\top}_{i}\BS_{t_m} + b_{i} \right)\mid\BS_t\right]+b_2
\end{eqnarray}

$\E\left[\varphi \left(\mathbf{w}^{\top}_{i}\BS_{t_m} + b_{i} \right)\mid\BS_t\right]$ is the risk-neutral price of a European arithmetic basket option that matures at $t_m$ with weight vector $\mathbf{w}_i,$ and strike $b_i.$  When the underlying asset follows a geometric Brownian motion (GBM) closed form solution for the arithmetic basket options are not available. We show a workaround in Section \ref{section6} , for the GBM case, for faster evaluation of $\E\left[\varphi \left(\mathbf{w}^{\top}_{i}\BS_{t_m} + b_{i} \right)\mid\BS_t\right].$

Summarizing, $\tilde{G}^{\beta_{t_m}}(\BS_{t_m})$ is the value of the payoff from a portfolio of arithmetic basket options that mature at $t_m.$ The structure of the options (weights and strikes) is determined by the outcome of hidden layer 1, while the amount of each option you need to hold in your portfolio is the outcome of the hidden layer 2. The value of the portfolio under the no-arbitrage and Markovian model assumption for any $t \in (t_{m-1}, t_m]$ is given by $\E\left[\tilde{G}^{\beta_{t_m}}(\BS_{t_m}) \mid \BS_t \right],$ and it replicates the corresponding value of the contingent claim $V_t$ at that point of time. As the weights are not a function of $t \in (t_{m-1},t_m)$ or the corresponding $\BS_t$ values, the portfolio is static between the monitoring dates. In a practical setting there is a risk if the Markovian assumption doesn't hold and the model parameters $\Theta$ change during the course of period. Therefore, it is more of semi-static hedging than the traditional static hedging.  

\subsection{Convergence}

While we have shown that when the value of the target contingent claim $V_{t_m}, \ m = 1,\ldots,M,$ is known, then between monitoring dates $(t_{m-1}, t_m]$ a static portfolio can be set up that replicates the value of contingent claim. We next show that RLNN would converge to the true price of the contingent claim at any time point, including $t_0.$ We still hold the assumptions that the underlying follows a Markovian process, the market is frictionless and there is no arbitrage. We additionally assume that  $\BS_{t_m} \in \I_d, \ m = 0,\ldots, M,$ almost surely, where $\I_d$ is some d-dimensional hypercube. \\

\begin{mydef6}\label{mainTheorem}
Under the assumptions stated above, for any $\epsilon >0$ 

\begin{equation}
\sup_{\BS_{t_{m-1}} \in \I_d}|V_{t_{m-1}}(\BS_{t_{m-1}}) - \tilde{V}_{t_{m-1}}(\BS_{t_{m-1}}) | < \epsilon, \ m = 1, \ldots, M+1,
\end{equation}

where $\tilde{V}_{t_{m-1}}$ is obtained by following the iteration defined by Equation \ref{optPortEqn} and Equation \ref{vIteration}.
\end{mydef6}

\emph{Proof:} We prove Theorem \ref{mainTheorem} by induction. 

At the maturity of the contingent claim the proof for  $V_{t_M} = \tilde{V}_{t_M},$ is trivial, as $\widehat{Q}_{t_M}$ is equal to 0 and the payoff $h$ is deterministic and known. 

Assuming,
\begin{equation*}
\sup_{\BS_{t_m} \in \I_d}|V_{t_m} (\BS_{t_m}) - \tilde{V}_{t_m}(\BS_{t_m})| < \frac{\epsilon}{2},
\end{equation*} and by the application of Lemma \ref{lemma1}, i.e.

\begin{equation*}
\sup_{\BS_{t_m} \in \I_d}|\tilde{V}_{t_m} (\BS_{t_m}) - \tilde{G}^{\beta_{t_m}}(\BS_{t_m} )| < \frac{\epsilon}{2},
\end{equation*}

we can then show with the use of triangular inequality that, 

\begin{eqnarray}\nonumber\label{triangResult}
|V_{t_m} (\BS_{t_m}) - \tilde{G}^{\beta_{t_m}}(\BS_{t_m}) | &\leq& |V_{t_m} (\BS_{t_m}) - \tilde{V}_{t_m}(\BS_{t_m})|+ |\tilde{V}_{t_m} (\BS_{t_m}) - \tilde{G}^{\beta_{t_m}}(\BS_{t_m} )|\\
 &<& \frac{\epsilon}{2}+\frac{\epsilon}{2} = \epsilon.
\end{eqnarray}

We next want to show that,

\begin{equation}\label{qConverge}
\sup_{\BS_{t_{m-1}} \in \I_d}|Q_{t_{m-1}} (\BS_{t_{m-1}}) - \widehat{Q}_{t_{m-1}}(\BS_{t_{m-1}}) | < \epsilon,
\end{equation}

where $ \widehat{Q}_{t_{m-1}}(\BS_{t_{m-1}}):=\E\left[\tilde{G}^{\beta_{t_m}}(\BS_{t_m} )\mid\BS_{t_{m-1}}\right],$ as defined in Equation \ref{optPortEqn}.

\begin{eqnarray} \nonumber
\abs{Q_{t_{m-1}} (\BS_{t_{m-1}}) - \widehat{Q}_{t_{m-1}}(\BS_{t_{m-1}})} &=& \abs{\E\left[V_{t_m}(\BS_{t_m})|\BS_{t_{m-1}}\right]-  \E\left[\tilde{G}^{\beta_{t_m}}(\BS_{t_m} )\mid\BS_{t_{m-1}}\right]}\\\nonumber
&=& \abs{\E\left[V_{t_m}(\BS_{t_m}) -\tilde{G}^{\beta_{t_m}}(\BS_{t_m} )\mid\BS_{t_{m-1}}\right]}\\\nonumber
&\leq&\E\left[\abs{V_{t_m}(\BS_{t_m}) -\tilde{G}^{\beta_{t_m}}(\BS_{t_m} )}\mid\BS_{t_{m-1}}\right]\\\nonumber
&<& \E\left[\epsilon \mid\BS_{t_{m-1}}\right] = \epsilon,
\end{eqnarray}

where the first equation is restating the definition, the second is the outcome of no arbitrage and linearity of expectation operator, the third line uses Jensen's inequality, while the final line is the outcome of substitution of the result in Equation \ref{triangResult}.

The following cases then arise:

\begin{itemize}
\item Case 1:  
\begin{flalign*} 
h(\BS_{t_{m-1}}) &> Q_{t_{m-1}} (\BS_{t_{m-1}}) > \widehat{Q}_{t_{m-1}}(\BS_{t_{m-1}}),&&
\end{flalign*}
or
\begin{flalign*} 
h(\BS_{t_{m-1}}) &> \widehat{Q}_{t_{m-1}}(\BS_{t_{m-1}})> Q_{t_{m-1}} (\BS_{t_{m-1}}) &&
\end{flalign*}

For both these sub-cases

 \begin{eqnarray}\nonumber
 \abs{V_{t_{m-1}}(\BS_{t_{m-1}})-\tilde{V}_{t_{m-1}}(\BS_{t_{m-1}})} &=& \abs{\max\left(h(\BS_{t_{m-1}}), Q_{t_{m-1}}(\BS_{t_{m-1}})\right)-\max(h(\BS_{t_{m-1}}), \widehat{Q}_{t_{m-1}}(\BS_{t_{m-1}}))}\\ \nonumber
 &=&\abs{h(\BS_{t_{m-1}}) - h(\BS_{t_{m-1}})} = 0 < \epsilon
 \end{eqnarray}

\item Case 2:
\begin{flalign*} 
h(\BS_{t_{m-1}}) &<Q_{t_{m-1}} (\BS_{t_{m-1}}) < \widehat{Q}_{t_{m-1}}(\BS_{t_{m-1}}),&&
\end{flalign*}
or
\begin{flalign*} 
h(\BS_{t_{m-1}}) &< \widehat{Q}_{t_{m-1}}(\BS_{t_{m-1}}) < Q_{t_{m-1}} (\BS_{t_{m-1}})&&
\end{flalign*}
 
 For both these sub-cases 
 \begin{eqnarray}\nonumber
 \abs{V_{t_{m-1}}(\BS_{t_{m-1}})-\tilde{V}_{t_{m-1}}(\BS_{t_{m-1}})} &=& \abs{\max\left(h(\BS_{t_{m-1}}), Q_{t_{m-1}}(\BS_{t_{m-1}})\right)-\max(h(\BS_{t_{m-1}}), \widehat{Q}_{t_{m-1}}(\BS_{t_{m-1}}))}\\ \nonumber
 &=&\abs{Q_{t_{m-1}}(\BS_{t_{m-1}}) -  \widehat{Q}_{t_{m-1}}(\BS_{t_{m-1}})} < \epsilon,
 \end{eqnarray}
 
 where the inequality in the second line is the outcome of Equation \ref{qConverge}.
 
 \item Case 3:
\begin{flalign*} 
Q_{t_{m-1}} (\BS_{t_{m-1}}) &< h(\BS_{t_{m-1}})  < \widehat{Q}_{t_{m-1}}(\BS_{t_{m-1}}),&&
\end{flalign*}

For this sub-case
\begin{flalign*} 
\abs{h(\BS_{t_{m-1}}) - \widehat{Q}_{t_{m-1}}(\BS_{t_{m-1}})} < \abs{Q_{t_{m-1}}(\BS_{t_{m-1}}) -  \widehat{Q}_{t_{m-1}}(\BS_{t_{m-1}})} 
\end{flalign*}
Therefore,
 \begin{eqnarray}\nonumber
 \abs{V_{t_{m-1}}(\BS_{t_{m-1}})-\tilde{V}_{t_{m-1}}(\BS_{t_{m-1}})} &=& \abs{\max\left(h(\BS_{t_{m-1}}), Q_{t_{m-1}}(\BS_{t_{m-1}})\right)-\max(h(\BS_{t_{m-1}}), \widehat{Q}_{t_{m-1}}(\BS_{t_{m-1}}))}\\ \nonumber
 &=&\abs{h(\BS_{t_{m-1}}) -  \widehat{Q}_{t_{m-1}}(\BS_{t_{m-1}})} <\abs{Q_{t_{m-1}}(\BS_{t_{m-1}}) -  \widehat{Q}_{t_{m-1}}(\BS_{t_{m-1}})} < \epsilon,
 \end{eqnarray}

\item Case 4:

\begin{flalign*} 
\widehat{Q}_{t_{m-1}}(\BS_{t_{m-1}}) &< h(\BS_{t_{m-1}}) < Q_{t_{m-1}} (\BS_{t_{m-1}})&&
\end{flalign*}
For this sub-case
\begin{flalign*} 
\abs{Q_{t_{m-1}}(\BS_{t_{m-1}}) - h(\BS_{t_{m-1}}) } < \abs{Q_{t_{m-1}}(\BS_{t_{m-1}}) -  \widehat{Q}_{t_{m-1}}(\BS_{t_{m-1}})} 
\end{flalign*}
Therefore,
 \begin{eqnarray}\nonumber
 \abs{V_{t_{m-1}}(\BS_{t_{m-1}})-\tilde{V}_{t_{m-1}}(\BS_{t_{m-1}})} &=& \abs{\max\left(h(\BS_{t_{m-1}}), Q_{t_{m-1}}(\BS_{t_{m-1}})\right)-\max(h(\BS_{t_{m-1}}), \widehat{Q}_{t_{m-1}}(\BS_{t_{m-1}}))}\\ \nonumber
 &=&\abs{Q_{t_{m-1}}(\BS_{t_{m-1}}) - h(\BS_{t_{m-1}})} <\abs{Q_{t_{m-1}}(\BS_{t_{m-1}}) -  \widehat{Q}_{t_{m-1}}(\BS_{t_{m-1}})} < \epsilon,
 \end{eqnarray}
\end{itemize}

With this we have shown that for all possible cases that might arise,
$\abs{V_{t_{m-1}}(\BS_{t_{m-1}})-\tilde{V}_{t_{m-1}}(\BS_{t_{m-1}})} < \epsilon.$ This ends the proof.

\section{Lower and Upper bounds}\label{section5}

Once the neural networks in the RLNN have been trained to approximate the value functions at each monitoring dates, the lower and upper bounds to the true price can be obtained in fairly straightforward manner as discussed below. An upper and lower bound estimate for the true price, rather than just relying on the direct price estimate from the RLNN method, i.e. $\tilde{V}_{t_0},$ is important as unlike linear regression models where the coefficients can be determined exactly, with neural networks typically it is difficult to directly infer whether the coefficients have converged after certain number of iterations of the stochastic gradient descent algorithm. A tight lower and upper bound value would then be an indicator of the quality of the RLNN price estimate.

\subsection{Lower Bound}

In order to obtain a lower bound to the true price, we first simulate a new set of $N_L$  paths following Equation \ref{dis}. The early stopping times for each path, $n =1,\ldots,N_L,$ are then determined as:

\begin{equation}
\tau(n) := \inf\left\lbrace t_m \in \{t_0,\ldots,t_M\}: h\left(\BS_{t_m}(n)\right) > \widehat{Q}_{t_m}\left(\BS_{t_m}(n)\right)\right\rbrace,
\end{equation}

where $\widehat{Q}_{t_m}\left(\BS_{t_m}(n)\right)$ is computed as Equation \ref{optPortEqn}, using the already obtained vector of free parameters $\beta_{t_{m+1}}$ from RLNN.  An unbiased estimate of the  lower bound is then obtained as

$$
\underline{\tilde{V}}_{t_0} = \frac{B_{t_0}}{N_L}\sum_{n = 1}^{N_L} \frac{h\left(\BS_{\tau(n)}\right)}{B_{\tau(n)}}.
$$

\subsection{Upper Bound}

 \cite{haugh} and  \cite{rogers} proposed the dual formulation for pricing Bermudan options. For an arbitrary adapted super-martingale process $\M_t,$ it follows that,
\begin{eqnarray}\nonumber \label{dual}
V_{t_0}(\BS_{t_0}) &=& \sup_{\tau }\E\left[\frac{h_{\tau}}{B_{\tau}}\right] \\ \nonumber
&=& \sup_{\tau}\E\left[\frac{h_{\tau}}{B_{\tau}} +\M_{\tau} - \M_{\tau}\right]\\ \nonumber
& \leq & \M_{t_0} + \sup_{\tau }\E\left[\frac{h_{\tau}}{B_{\tau}} - \M_{\tau}\right]\\ \nonumber
& \leq & \M_{t_0} + \E\left[\max_{t}\left( \frac{h_{t}}{B_{t}} - \M_{t} \right)\right],
 \end{eqnarray}

which gives us the upper bound of the option price $V_{t_0}(\BS_{t_0}).$ Thus, the dual problem is to minimize the upper bound with respect to all adapted super-martingale processes, i.e.,

\begin{equation}
\overline{V}_{t_0}(\BS_{t_0})=\inf_{\M\in\Pi}\left(\M_{t_0}+\E\left[\max_{t}\left(\frac{h_{t}}{B_{t}}-\M_{t}\right)\right]\right),
\end{equation}

where $\Pi$ is the set of all adapted super-martingale processes. 

With, $\M_{t_0} = 0$, we construct a martingale process as:

\begin{equation}\label{martingale}
\M_{t_m}(n) = \sum_{i=0}^{m-1} \left[\frac{\tilde{G}^{\beta_{t_{i+1}}}\left(\mathbf{S}_{t_{i+1}}(n)\right)}{B_{t_{i+1}}} - \frac{\widehat{Q}_{t_i}\left(\BS_{t_i}(n)\right)}{B_{t_i}}\right]
\end{equation}

The upper bound, $\overline{V}_{t_0},$  is then given by

\begin{eqnarray}\nonumber
\overline{V}_{t_0}(\BS_{t_0}) &=& \E\left[\max_{t} \left(\frac{h_t}{B_t} - \M_t\right)\right]\\\label{dual3}
&=&\frac{1}{N_L} \sum_{n=1}^{N_L} \max_{t_m} \left(\frac{h(\BS_{t_m}(n))}{B_{t_m}} - \M_{t_m}(n)\right),\,t_m\in[t_0,\ldots,t_M] 
\end{eqnarray}

The important point here, as was also pointed out by \cite{glasserman2004simulation}, is that the martingale terms given by Equation \ref{martingale} are available at almost no extra cost. This is possible because we assume that $\tilde{G}^{\beta_{t_{m+1}}}$ has a closed form conditional expectation, which by construction is equal to $\widehat{Q}_{t_m}.$   In contrast, for \emph{regress now} schemes, in order to obtain the optimal martingale, one has to resort to sub-simulations which makes them computationally quite expensive. \\

\begin{mydef}
From a practical perspective the lower bound and upper bounds can be interpreted as following. For the holder of an early exercise option, the difference between the true price and the lower bound, gives the risk-neutral expectation of the loss he incurs when he exercises his option following the policy obtained using the RLNN. On the other hand, the difference between the upper bound and the true price is an indicator of the expected maximum loss the writer of the option has to bear for imperfect hedging. Equation \ref{martingale} can be seen as the rolling over of the semi-static hedge portfolio, where at $t_i$ an amount $\widehat{Q}_{t_i}(\BS_{t_i}(n)$ is used to set up a portfolio of European options that result in a payoff of $\tilde{G}^{\beta_{t_{i+1}}}(\BS_{t_{i+1}}(n))$ at $t_{i+1}.$ If the holder of the target option does not decide to exercise his option at $t_{i+1}$ the proceeds from the payoff of the static hedge portfolio would be greater than or just enough to set up the static hedge, a portfolio with options maturing at $t_{i+2},$ i.e. 

$$
\tilde{G}^{\beta_{t_{i+1}}}(\BS_{t_{i+1}}(n)) \geq \widehat{Q}_{t_{i+1}}(\BS_{t_{i+1}}(n).
$$

In case the holder of the target option decides to exercise at $t_{i+1},$  $\tilde{G}^{\beta_{t_{i+1}}}(\BS_{t_{i+1}}(n))$ should ideally be greater than equal to $h(\BS_{t_{i+1}}), $ or otherwise the writer of the target option suffers a loss. Therefore, Equation \ref{dual3} gives the expectation of the worst-case hedging loss for the writer of the target option along each simulated risk-neutral scenario.
\end{mydef}

On the practical side we show through numerical examples that  RLNN can be used to obtain tight upper and lower bounds without the need for any sub-simulations.

\section{Numerical Examples}\label{section6}

In this section we illustrate the performance of the RLNN by pricing path dependent and path independent contingent claims on single or several assets. We through experiments illustrate the two major themes covered in this paper. First, we look at the pricing of Bermudan options for different number of underliers and different payoff structures. Specifically, we look at the pricing of the following Bermudan options, put on a single asset, arithmetic basket put option on five assets, and max call option on two, three and five assets respectively. We also construct the upper and lower bounds for these examples using the methodology discussed in Section \ref{section5}. 
Second, through numerical examples, we compare the performance of dynamic hedging against the semi-static hedging technique discussed here, for a single asset European and Barrier option case.

For the examples considered here, we assume that the $d$ underlying assets follow a multivariate geometric Brownian motion, i.e.

\begin{equation}
\frac{dS_t^{\delta}}{S_t^{\delta}} = (r-q_{\delta})dt + \sigma_{\delta} dW_t^{\delta}, \, \delta =1,\ldots,d,
 \end{equation}

where $r$ is the constant risk-free rate, $q_{\delta}$ is the continuous dividend rate for the $\delta$-th asset. $W_t^{\delta}$ is the standard Brownian motion and the instantaneous correlation coefficient between $W_t^i$ and $W_t^j$ is $\rho_{ij}.$

\begin{table}
\center
\begin{tabular}{|l|}\hline\\
\underline{Set I}\\
\\
 $S^{1}_{t_0}=40,$ $K=40,$ $r = 0.06,$ $\sigma=0.2,$ $T=1,$ $M=10.$\\
 \\
 \underline{Set II}\\
 \\

 $S^{\delta}_{t_0}=1,$ $K=1,$ $r = 0.05,$ $q_{\delta}=0,$  $T=1,$ $M=10.$ \\
 \\
 $\sigma= \left[ 0.518, 0.648, 0.623, 0.570, 0.530\right]^{\top}$\\
 \\
 $\rho:=\begin{pmatrix}
 1.00 &0.79 &0.82 &0.91 &0.84\\
0.79 &1.00 &0.73 &0.80 &0.76\\
0.82 &0.73 &1.00 &0.77 &0.72\\
0.91 &0.80 &0.77 &1.00 &0.90\\
0.84 &0.76 &0.72 &0.90 &1.00
 \end{pmatrix}$
 \\
 \\
\underline{Set III} \\
\\
$S^{\delta}_{t_0}=100,$ $K=100,$ $r = 0.05,$ $q_{\delta}=0.1,$ $\sigma_{\delta}=0.2,$ $\rho_{ij}=0.0,$ $T=3,$ $M=9.$\\ 
\\
\underline{Set IV} \\
\\
$S^{1}_{t_0}=1,$ $K=1,$ $r = 0.1,$  $\sigma_{1}=0.3,$  $T=1.$ \\ 
\\
\underline{Set V} \\
\\
$S^{1}_{t_0}=1,$ $K=1,$ $r = 0.1,$  $\sigma_{1}=0.3,$ $T=0.2,$ $M=5.$\\ 
\\

\\\hline

\end{tabular}
\caption{ Parameter values used in the examples. }\label{params}
\end{table}

The implementation of all the examples considered are in python, where we use Keras with Tensorflow backend for defining and training the neural networks. 

\subsection{Choice of hyper-parameters and other considerations}

The choice of appropriate hyper-parameters for training the neural network can affect the rate of convergence and  therefore should be carefully picked. For the experiments we conducted we made the following choices. 

\begin{itemize}
\item We use Adam \cite{adam} with initial leaning rate as $10^{-3},$ as the optimizer for the weights update in the mini-batch gradient ascent algorithm. 
\item The batch size is chosen as one tenth of the total training points. 
\item For training the neural network corresponding to the monitoring date $t_M=T,$ we initialize the weights and biases randomly with uniform
random variables.
\item As $\tilde{V}_{t_{m-1}} \approx \tilde{V}_{t_m}$ we transfer the final weights obtained from training the network for monitoring date $t_m$ as the initial weights for training the network at $t_{m-1}.$  
\item In order to avoid over-fitting, we divide the training points into training set and validation set, in the ratio 0.7 to 0.3 respectively and use the mean squared error of the validation set as the early-stopping criteria with a patience of 6 epochs. 
\item We normalize the initial asset price to 1 and appropriately adjust the strike. This restricts the domain in which the network needs to be trained, something we found especially beneficial while training the network for max options. 
\item We use 50,000 training points generated using the GBM process under the risk-neutral measure. 
\end{itemize}

For obtaining the upper and lower bound estimates we generate a fresh set of 200,000 scenarios for each of the example considered.

\subsection{Single Asset Bermudan}

We first consider pricing of a Bermudan put on a single asset. The parameter set for the experiment corresponds to Set 1 in Table \ref{params}. 
Following the algorithm \ref{alg1}, the input to the network at each monitoring date $t_m$ is one dimensional $S_{t_m}.$ The output from the network at $t_m$ is therefore:
\begin{equation*}
\tilde{G}^{\beta_{t_m}}(S_{t_m}) = \sum_{i = 1}^p \omega_{i} \varphi \left({w}_{i}S_{t_m} + b_{i} \right)+b_2
\end{equation*}  

In order to compute the continuation value $\widehat{Q}_{t_{m-1}},$ we need closed form expression for Equation \ref{optPortEqn}, i.e. we need to evaluate,

\begin{eqnarray}\nonumber
\widehat{Q}_{t_{m-1}}(S_{t_{m-1}}) &=& \E\left[\tilde{G}^{\beta_{t_m}}(S_{t_m}) \mid S_{t_{m-1}}\right]\\\nonumber
&=& \E\left[\sum_{i = 1}^p \omega_{i} \varphi \left({w}_{i}S_{t_m} + b_{i} \right)+b_2 \mid S_{t_{m-1}}\right]\\ \nonumber
&=& \sum_{i = 1}^p \omega_{i} \E\left[\varphi \left(w_{i}S_{t_m} + b_{i} \right) \mid S_{t_{m-1}}\right]+b_2 
\end{eqnarray}

Depending upon the signs of $w_i,$ and $b_i$ $\E\left[\varphi \left(w_{i}S_{t_m} + b_{i} \right)\mid S_{t_{m-1}}\right]$ is the value of call option, put option, or a forward contract. These possible cases might then arise:

\begin{enumerate}
\item[Case 1] $w_i$ and $b_i$ are both greater than 0. Then 
\begin{flalign*} 
\E\left[\max\left(w_iS_{t_m}+b_i,0\right)|S_t\right] &= \E\left[\left(w_iS_{t_m}+b_i\right)|S_{t_{m-1}}\right],&&
\end{flalign*}
which is the price of a forward contract. 
\item[Case 2] $w_i >0 $ and $b_i <0,$ then 

\begin{flalign*}
\E\left[\max\left(w_iS_{t_m}+b_i,0\right)\mid S_t\right] &= w_i\E\left[\max\left(S_{t_m}+\frac{b_i}{w_i},0\right)\mid S_{t_{m-1}}\right],&&
\end{flalign*}
which is the time $t_{m-1}$ value of a European call option, with strike $-\frac{b_i}{w_i},$ that expires at $t_m.$
\item[Case 3] $w_i  <0 $ and $b_i  >0,$ then
 
\begin{flalign*}
\E\left[\max\left(w_iS_{t_m}+b_i,0\right)\mid S_t\right] &= -w_i\E\left[\max\left(-\frac{b_i}{w_i} - S_{t_m},0\right)\mid S_{t_{m-1}}\right],&&
\end{flalign*}
which is the time $t_{m-1}$ value of a European put option, with strike $-\frac{b_i}{w_i},$ that expires at $t_m.$
\item[Case 4] $w_i <0$ and $b_i <0,$ then 

\begin{flalign*}
\E\left[\max\left(w_iS_{t_m}+b_i,0\right)\mid S_t\right] &=0,&&
\end{flalign*}
\end{enumerate}

From a practical point of view, when the asset follows a GBM process, the implementation can be greatly simplified by taking the log of the asset price as the input to the neural network. The continuation value is then:

\begin{eqnarray}\nonumber
\widehat{Q}_{t_{m-1}}(S_{t_{m-1}}) &=& \E\left[\tilde{G}^{\beta_{t_m}}(\log(S_{t_m})) \mid S_{t_{m-1}}\right]\\\label{logNetwork}
&=& \sum_{i = 1}^p \omega_{i} \E\left[\varphi \left(w_{i}\log(S_{t_m}) + b_{i} \right) \mid S_{t_{m-1}}\right]+b_2,
\end{eqnarray}

In \ref{Ax} we give the solution for Equation \ref{logNetwork}, in a more generic  multi-asset context, where $\BS_t$ follows a $d-$dimensional Geometric Brownian motion, and an element-wise log transform of the asset prices is used as the input to the neural network. 

\begin{figure}
 \centerline{\includegraphics[width=0.7\textwidth]{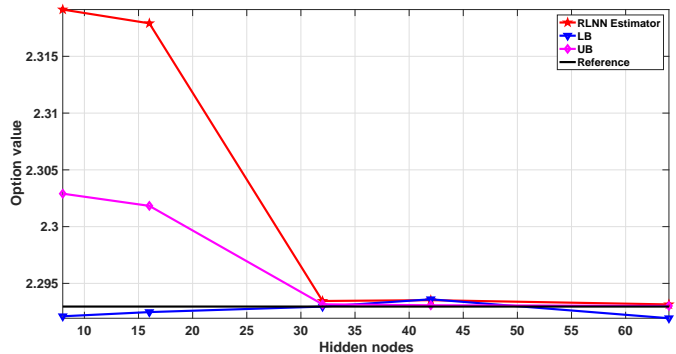}}
  \label{fig2}
  \caption{The upper and lower bound values of a Bermudan call on a single asset, when an increasing number of hidden units are used in the first layer of RLNN. The parameters for the model and the option are taken from Set I in Table \ref{params}. The reference value is obtained using COS method and is equal to 2.2929. Also plotted are the corresponding direct RLNN estimated values, i.e. $\tilde{V}_{t_0}.$ }

\end{figure}

Table \ref{res1} gives the price of Bermudan put for different initial asset prices. The standard errors are obtained using the results from 30 independent runs. The reference value is obtained using the COS method \cite{COS}. The convergence of the results to the true price with increasing number of hidden units, or short maturity options used in the static hedge portfolio, is illustrated in Figure \ref{fig2}. We observe that the upper bound values have much lower standard errors in comparison to the lower bound estimates, something that we consistently observe in the other examples considered here as well.

\begin{table}\centering
\scalebox{0.7}
{

                   \begin{tabular}{|c|c|c|c|c|c|}
                   \hline 
                   $S_0$ & RLNN  & RLNN & RLNN & RLNN &  COS\\
                  
                     & Direct est. (s.e.) & Lower Bound. (s.e.) & Upper Bound (s.e.) & 95\% CI &  \\ 
                   \hline
                  
                   36 &  4.4429 &  4.4439  &  4.4427 & [4.4422,	 4.4427] & 4.4425\\ 
                  
                     & (0.0007) & (0.004) & (0.00004) &   &   \\ 
                   \hline 
                   40 & 2.2934&  2.2935& 2.2930 & [ 2.2919,   2.2930] ]&2.2929\\ 
                 
                     & (0.0007) & (0.004) & ( 0.00001) &   &   \\ 
                   \hline 
                   44  & 1.0989&  1.0979 &  1.0987 & [ 1.0970,   1.0987] & 1.0984 \\ 
                 
                     & (0.0008) & (0.002) & (0.00002) &   &   \\ 
                   \hline
                   \end{tabular}
 }
 \caption{The RLNN direct estimator, upper, and lower bound values of a Bermudan call on a single asset when 32 hidden units are used in the first layer of the neural networks in RLNN. For obtaining the lower and upper bound values we use 200,000 independent paths. The parameters for the model and the option are taken from Set I in Table \ref{params}. The reference values are obtained using the COS method. }  \label{res1}

\end{table}

\subsection{Arithmetic Basket Option}

The weighted basket put option payoff reads as,

$$
h(\BS_T) = \max\left(K-\sum_{i = 1}^{d} w_i S_T^i ,0\right).
$$

The model parameters are the same ones as used by  \cite{reisinger} to approximate the DAX index, and corresponds to Set 2 in Table \ref{params}. The weights for the basket are taken as
$$
 w = [0.381, 0.065, 0.057, 0.270, 0.227]^{\top}.
$$

We consider a Bermudan put option that expires at $T = 1$ year and has an early exercise option on $M=10$ equally spaced time points between $t = 0$ and $t=T.$ 

When the input to the neural network at $t_m$ are the asset prices $\BS_{t_m},$ the output of the network is a portfolio of European arithmetic basket options that expires at $t_m.$ While fast numerical approximations and semi-analytical expressions for computing the price of European arithmetic basket options are available, the problem can be greatly simplified --- when the  underlying assets follow the GBM process--- if the log asset prices are taken as the input to the neural network. With log asset prices as the input, the output of the network at $t_m$ translates to the payoff of a portfolio of European geometric basket options that mature at $t_m.$ In order to compute the continuation value, one needs to compute the following:

\begin{eqnarray}\nonumber
\widehat{Q}_{t_{m-1}}(S_{t_{m-1}}) &=& \E\left[\tilde{G}^{\beta_{t_m}}(\log\left(\BS_{t_m}\right)) \mid \BS_{t_{m-1}} \right] \\\nonumber
 &=& \E\left[\sum_{i = 1}^p \omega_{i} \varphi \left(\mathbf{w}^{\top}_{i}\log\left(\BS_{t_m}\right) + b_{i} \right)+b_2\mid \BS_{t_{m-1}}\right]\\ \nonumber
&=& \sum_{i = 1}^p \omega_{i} \E\left[\varphi \left(\mathbf{w}^{\top}_{i}\log\left(\BS_{t_m}\right) + b_{i} \right)\mid\BS_{t_{m-1}}\right]+b_2
\end{eqnarray}

In \ref{Ax} we provide the expression for $ \E\left[\varphi \left(\mathbf{w}^{\top}_{i}\log\left(\BS_{t_m}\right) + b_{i} \right)\mid\BS_{t_{m-1}}\right],$ which is required for evaluation of $\widehat{Q}_{t_{m-1}}.$

\begin{figure}
 \centerline{\includegraphics[width=0.7\textwidth]{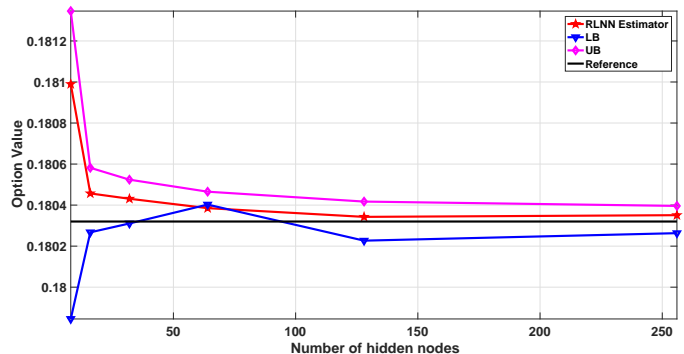}}
  \label{fig3}
  \caption{The upper and lower bound values for a Bermudan arithmetic basket put option, when an increasing number of hidden units are used in the first layer of RLNN. The parameters for the model and the option are taken from Set II in Table \ref{params}. The reference value is obtained using SGBM method and is equal to 0.1803. Also plotted are the corresponding direct RLNN estimated values, i.e. $\tilde{V}_{t_0}.$ }

\end{figure}

Table \ref{res2} provides the lower and upper bound values for the Bermudan arithmetic basket put option obtained using the RLNN method. The reported standard errors are obtained using results from 30 independent runs. The reference values are obtained using the SGBM method \cite{sgbm}. We again notice that the standard errors for the duality based upper bound values are order of magnitude smaller than others. Figure \ref{fig3} illustrates the convergence of the results for the arithmetic basket case when an increasing number of hidden units are used. 

\begin{table}\centering
\scalebox{0.7}
{

                   \begin{tabular}{|c|c|c|c|c|c|}
                   \hline 
                   $S_0$ & RLNN  & RLNN & RLNN & RLNN &  SGBM\\
                  
                     & Direct est. (s.e.) & Lower Bound. (s.e.) & Upper Bound (s.e.) & 95\% CI &  \\ 
                   \hline
                  
                   0.9 &  0.2222 &  0.2221  &  0.2223& [0.2219, 0.2223] & 0.2220\\ 
                  
                     & (0.0001) & (0.0005) & (0.00001) &   &   \\ 
                   \hline 
                   1 & 0.1804&  0.1803 & 0.1805 & [0.1802, 0.1805]&0.1803 \\ 
                 
                     & (0.00009) & (0.0003) & (0.00001) &   &   \\ 
                   \hline 
                   1.1  & 0.1464&  0.1464&  0.1464& [0.1463, 0.1464] & 0.1463\\ 
                 
                     & (0.0001) & (0.0003) & (0.00001) &   &   \\ 
                   \hline
                   \end{tabular}
 }
 \caption{The RLNN direct estimator, upper, and lower bound values of a Bermudan arithmetic basket put option on five assets when 64 hidden units are used in the RLNN method. For obtaining the lower and upper bound values we use 200,000 independent paths. The parameters for the model and the option are taken from Set II in Table \ref{params}. The reference values are obtained using the SGBM method \cite{sgbm}. }  \label{res2}

\end{table}

\subsection{Max Option}

We next consider a Bermudan max call  option, whose payoff at $t$ is given by:
$$
h(\BS_t) = \max\left(\max\left(S_t^1,\ldots,S_t^d\right)-K,0\right)
$$

The parameters for this case correspond to Set III in the Table \ref{params}. Similar to the case of Bermudan arithmetic basket option, one can either use the asset prices as inputs to the neural network, in which case the output would correspond to a portfolio of short maturity arithmetic basket options. When the underlying assets follow a GBM process, it might be more convenient to use the log asset prices as the input to the neural network, in which case the output of the network corresponds to a static hedge with a portfolio of geometric basket options. 

The computation of the continuation value remains the same as that for the Bermudan arithmetic basket case. Unlike the linear regression-based models, with RLNN we do not have to make payoff specific choices for basis functions, which greatly simplifies the implementation and maintenance of the pricing library. 

\begin{figure}
 \centerline{\includegraphics[width=0.7\textwidth]{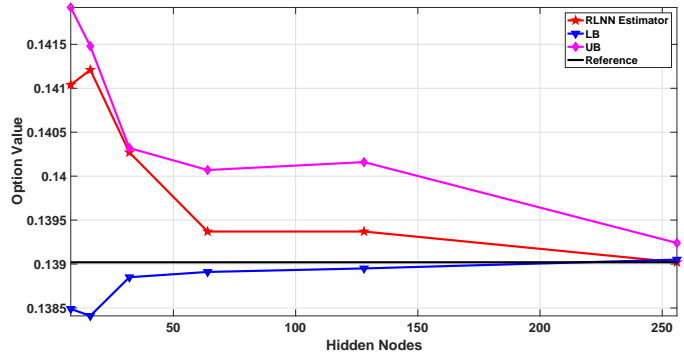}}
  \label{fig4}
  \caption{The upper and lower bound values for a Bermudan call on the maximum of two assets when an increasing number of hidden units are used in the first layer of RLNN. The parameters for the model and the option are taken from Set III in Table \ref{params}. The reference value is obtained from a binomial tree and is equal to 13.902. Also plotted are the corresponding direct RLNN estimator values, i.e. $\tilde{V}_{t_0}.$ }

\end{figure}

Table \ref{res3} provides the lower and upper bounds for the Bermudan max option on 2, 3 and 5 assets and compares the results with benchmark values reported in \cite{Cao} and \cite{Andersen}. The convergence of the results to the reference value with increasing number of hidden units  is illustrated in Figure \ref{fig4}.

\begin{table}\centering
\scalebox{0.7}
{

                   \begin{tabular}{|c|c|c|c|c|c|c|}
                   \hline 
                   $S_0$ & RLNN  & RLNN & RLNN & RLNN & Literature & Binomial\\ 
                  
                     & Direct est. (s.e.) & Lower Bound. (s.e.) & Upper Bound (s.e.) & 95\% CI & 95\% CI& \\ 
                   \hline
                    \multicolumn{7}{l}{d=2 assets:}\\
                   \hline 
                   90 &  8.078 &  8.071 &  8.086& [8.062,   8.086] & [8.053, 8.082] &8.075\\ 
                  
                     & (0.016) & (0.025) & (0.0003) &   &   &\\ 
                   \hline 
                   100 & 13.902&  13.905 & 13.924 & [ 13.894,   13.924] & [13.892, 13.934]&13.902\\ 
                 
                     & (0.017) & (0.028) & (0.0004) &   &  & \\ 
                   \hline 
                   110 & 21.346&  21.352 &  21.353 & [ 21.341,   21.353] & [21.316, 21.359]&21.345 \\ 
                 
                     & (0.010) & (0.029) & (0.0002) &   &   &\\ 
                   \hline
                    \multicolumn{7}{l}{d=3 assets:}\\
                   \hline 
                   90 & 11.282 & 11.295 & 11.303 & [11.287,   11.303] & [11.265,  11.308] &11.29\\ 
                  
                     & (0.017) & (0.030) & (0.001) &   &  & \\ 
                   \hline 
                   100 & 18.702 & 18.688 & 18.715 & [18.677,  18.715] & [18.661,  18.728]&18.69\\ 
                 
                     & (0.022) & (0.025) & (0.001) &   &   &\\ 
                   \hline 
                   110 & 27.572 & 27.554 & 27.592 & [27.545,  27.592] & [27.512,  27.663]&27.58 \\ 
                 
                     & (0.021) & (0.023) & (0.001) &   &  & \\ 
                   \hline 
                   \multicolumn{7}{l}{d=5 assets:}\\
                   \hline 
                   90 & 16.680 & 16.636 & 16.743 & [16.624,  16.744] & [16.620,  16.653]& \\ 
                  
                     & (0.063) & (0.044) & (0.003) &   &  & \\ 
                   \hline 
                   100 & 26.177 & 26.141 & 26.268 & [26.125,  26.270] & [26.115,  26.164]&\\ 
                 
                     & (0.062) & (0.034) & (0.002) &   &  & \\ 
                   \hline 
                   110 & 36.815 & 36.760 & 36.909 & [36.744,  36.909] & [36.710,  36.798]& \\ 
                 
                     & (0.042) & (0.045) & (0.003) &   &  & \\ 
                   \hline 
                   \end{tabular} 

}
\caption{\footnotesize Bermudan option values for call on maximum of 2, 3 and 5 assets, with parameter values taken from Set III, in Table \ref{params}. For the two asset case we use 256 hidden units, for three assets 512 hidden units, while for five assets we use 1024 hidden units. The reference confidence interval for the two and three asset case are taken from \cite{Andersen}, and for the five asset case from  \cite{Cao}. }\label{res3}

\end{table}

\subsection{Performance of the static hedge}

Here we consider the problem faced by the writer of a European vanilla put option with a maturity of one year, and a down-and-out call barrier option with maturity of 0.2 years. In the first case the writer wants to hold a short position on the European put option for a month, after which the option position is closed. The writer has the option to hedge the risk using either the dynamic delta hedging strategy, or using the static hedge portfolio of  short maturity vanilla options constructed using the RLNN method.  The dynamic hedging strategy involves daily rebalancing, in our example we use 25 equally distributed points between 0 to 1 month (i.e. $\frac{1}{12}$ year fraction). 
The second case involves the writer of a discretely monitored down-and-out-call barrier option, with a maturity of 60 days and 5 equally spaced monitoring dates, who wants to hold a short position for 12 days. He has a choice of hedging his risks using the static hedge portfolio constructed using RLNN or through dynamic delta hedging with daily rebalancing. 

The comparison for the above two cases here are based on Monte Carlo simulations, with 50000 paths, where the paths are generated following the Black Scholes model under the risk-neutral measure. In the limiting case both dynamic and static hedging results should work perfectly. However, we here consider the performance when only a fixed number of short maturity options are used for the static hedge, and the portfolio is rebalanced on finite time points for the dynamic hedge. Table \ref{tab:BS_Hedging_summary} outlines the results of the experiment for the European put option case. We see that the dynamic and static hedge strategies have similar performance, although the static hedge gives better results when it comes to VaR and conditional VaR. Also evident is that the performance of the static hedge improves when an increasing number of short maturity options are held in the static hedge portfolio. There, when the writer of the option is particularly concerned about avoiding large losses, he should prefer the static strategy.

\begin{table}[h]
\caption{Hedge Performance of European Put Options} 
\centering 
\hspace*{-2em}
\scalebox{0.7}
{
\begin{tabular}{c c c cccc}\\ 
\hline\hline 
 & & & \multicolumn{3}{c}{Moneyness (K/S)} \\ [0.5ex]
  \raisebox{1.5ex}{Hedging Error} & \raisebox{1.5ex}{Hedge Type}  & \raisebox{1.5ex}{Options Count}  & 0.5 & 1 & 1.5 
\\ [0.5ex]
\hline 
\\

  &   & 10 & -9e-06 & -4e-06& -1e-05  \\[1ex]
  & Static & 25 & -5e-07 & 1e-06 & 3e-05  \\[1ex]
\textbf{Mean} &  & 50 & 1e-07 & -2e-06 & 7e-06  \\[1ex]
\\
& Dynamic & NA & 5e-07 & -7e-06 & 1e-05 \\[3ex]
\hline
\\


  &   & 10 & 2e-05 & 0.0010 & -0.0090  \\[1ex]
  & Static & 25 & 1e-05 & 0.0002 & 0.0025  \\[1ex]
\textbf{Standard Deviation} &  & 50 & 1e-05 & 0.0002 & 0.0003  \\[1ex]
\\
& Dynamic & NA & 3e-05 & 0.0013 & 0.0011  \\[3ex]
\hline
\\

  &   & 10 & 5e-05 & 0.0017 & 0.0013  \\[1ex]
  & Static & 25 & 1e-05 & 0.0004 & 0.0036 \\[1ex]
\textbf{VaR(95 \%) } &  & 50 & 2e-05 & 0.0004 & 0.0004  \\[1ex]
\\
& Dynamic & NA & 6e-05 & 0.0026 & 0.0021 \\[3ex]
\hline
\\

  &   & 10 & 7e-05 & 0.0031 & 0.0033  \\[1ex]
  & Static & 25 & 2e-05 & 0.0006 & 0.0079  \\[1ex]
\textbf{CVaR (95\%)} &  & 50 & 3e-05 & 0.0007 & 0.0010  \\[1ex]
\\
& Dynamic & NA & 0.0001 & 0.0033 & 0.0027  \\[3ex]
\hline
%

\end{tabular}
}
\caption{\footnotesize This table compares the hedging performance of the dynamic and static hedging (constructed using the RLNN) strategies for a European Vanilla put option for a duration of one month. The results are reported for varying levels of moneyness ($K/S$). The parameters for the model and the barrier are taken from Set IV in Table \ref{params}.
}
\label{tab:BS_Hedging_summary}
\end{table}

Table \ref{tab:Barrier_Hedging_summary} outlines the results of the experiment for the discretely monitored down-and-out call barrier case. We see that the dynamic and static hedge strategies have similar performance for lower barrier values, but with an increasing barrier level the static hedge seems to have slight advantage over dynamic hedging. The static hedge, in most cases, gives better results when it comes to VaR and conditional VaR. As expected, the performance of the static hedge improves when an increasing number of short maturity options are held in the portfolio. 
\begin{table}[h]
\caption{Hedge Performance of Down and Out Barrier Call Options} 
\centering 
\hspace*{-1em}
\scalebox{0.7}
{
\begin{tabular}{c c c cccc}\\ 
\hline\hline 
 & & & \multicolumn{4}{c}{Barrier} \\ [0.5ex]
  \raisebox{1.5ex}{Hedging Error} & \raisebox{1.5ex}{Hedge Type}  & \raisebox{1.5ex}{Options Count}  & 0.91 & 0.93 & 0.95 & 0.97
\\ [0.5ex]
\hline 
\\

  &   & 5 & -0.0001 & -4e-5 & -1e-5 & -2e-5 \\[1ex]
  & Static & 10 &  -0.0002 & 1.6e-5 & -1e-5 & 5e-5 \\[1ex]
\textbf{Mean} &  & 20 & -0.0001 & -6e-5 & -8e-5 & 8e-5 \\[1ex]
\\
& Dynamic & NA & -0.0001 & -0.0002 & -0.0001 & 0.0008 \\[3ex]
\hline
\\


  &   & 5 & 0.0028 & 0.0037 & 0.0038 & 0.0042 \\[1ex]
  & Static & 10 & 0.0036 & 0.0046 & 0.0038 & 0.0042 \\[1ex]
\textbf{Standard Deviation} &  & 20 & 0.0024 & 0.0026 & 0.0038 & 0.0042 \\[1ex]
\\
& Dynamic & NA & 0.0028 & 0.0043 & 0.0076 & 0.0132 \\[3ex]
\hline
\\

  &   & 5 & 0.0062 & 0.0062 & 0.0077 & 0.0090 \\[1ex]
  & Static & 10 & 0.0076 & 0.0083 & 0.0077 & 0.0089 \\[1ex]
\textbf{VaR(95\%)} &  & 20 & 0.0049 & 0.0048 & 0.0076 & 0.0089 \\[1ex]
\\
& Dynamic & NA & 0.0048 & 0.0093 & 0.0168 & 0.0249 \\[3ex]
\hline
\\

  &   & 5 & 0.0100 & 0.0114 & 0.0106 & 0.0129 \\[1ex]
  & Static & 10 & 0.0122 & 0.0128 & 0.0106 & 0.0127 \\[1ex]
\textbf{CVaR(95\%)} &  & 20 & 0.0088 & 0.0091 & 0.0106 & 0.0128\\[1ex]
\\
& Dynamic & NA & 0.0085 & 0.0153 & 0.0263 & 0.0461 \\[3ex]
\hline

%

\end{tabular}
}
\caption{\footnotesize This table compares the performance of dynamic and static hedging strategies for a 60-days down-and-out call barrier options at varied levels of barrier for a duration of 12 days. The parameters for the model and the barrier are taken from Set V in Table \ref{params}.
}

\label{tab:Barrier_Hedging_summary}
\end{table}

  \section{Conclusion}\label{section7}

In this paper we have developed a fully interpretable neural network based pricing method for path dependent high dimensional contingent claims. The main theoretical contribution is that we show, under the Markovian and the no-arbitrage assumption, any contingent claim ---with continuous payoffs--- can be semi-statically hedged using a finite number of shorter maturity options. The structure and the quantity of the options to be held at different monitoring dates in the static hedge portfolio are the outcomes of the RLNN method.  We  show that using RLNN one can obtain lower and upper bounds to the true value of the claim and the quality of the bounds is demonstrated through numerical examples. An important contribution here is that, with RLNN one can obtain a duality based upper bound without the need for performing sub-simulations. Through numerical examples we show the accuracy of the price and bounds obtained using the RLNN method for high dimensional path dependent options with different payoff structures. Through simulations we demonstrate that the static hedge approximation performs as effectively or better than delta hedging with daily rebalancing.  We see the analysis of the performance of the static hedge under more realistic conditions as an important extension to this line of research.

\bibliographystyle{abbrvnat}
\bibliography{references}

\begin{thebibliography}{35}
\providecommand{\natexlab}[1]{#1}
\providecommand{\url}[1]{\texttt{#1}}
\expandafter\ifx\csname urlstyle\endcsname\relax
  \providecommand{\doi}[1]{doi: #1}\else
  \providecommand{\doi}{doi: \begingroup \urlstyle{rm}\Url}\fi

\bibitem[Ackerer et~al.(2019)Ackerer, Tagasovska, and Vatter]{ackerer2019deep}
D.~Ackerer, N.~Tagasovska, and T.~Vatter.
\newblock Deep smoothing of the implied volatility surface.
\newblock \emph{arXiv preprint arXiv:1906.05065}, 2019.

\bibitem[Andersen and Broadie(2004)]{Andersen}
L.~Andersen and M.~Broadie.
\newblock Primal-dual simulation algorithm for pricing multidimensional
  american options.
\newblock \emph{Management Science}, 50\penalty0 (9):\penalty0 1222--1234,
  2004.

\bibitem[Balata and Palczewski(2017)]{balata2017}
A.~Balata and J.~Palczewski.
\newblock Regress-later monte carlo for optimal control of markov processes.
\newblock \emph{arXiv preprint arXiv:1712.09705}, 2017.

\bibitem[Bayer et~al.(2019)Bayer, Horvath, Muguruza, Stemper, and
  Tomas]{bayer2019deep}
C.~Bayer, B.~Horvath, A.~Muguruza, B.~Stemper, and M.~Tomas.
\newblock On deep calibration of (rough) stochastic volatility models.
\newblock \emph{arXiv preprint arXiv:1908.08806}, 2019.

\bibitem[Becker et~al.(2019)Becker, Cheridito, and Jentzen]{becker}
S.~Becker, P.~Cheridito, and A.~Jentzen.
\newblock Deep optimal stopping.
\newblock \emph{Journal of Machine Learning Research}, 20\penalty0
  (74):\penalty0 1--25, 2019.

\bibitem[Bender and Schweizer(2019)]{bender}
C.~Bender and N.~Schweizer.
\newblock Regression anytime'with brute-force svd truncation.
\newblock \emph{arXiv preprint arXiv:1908.08264}, 2019.

\bibitem[Beutner et~al.(2013)Beutner, Pelsser, and Schweizer]{beutner2013fast}
E.~Beutner, A.~Pelsser, and J.~Schweizer.
\newblock Fast convergence of regress-later estimates in least squares monte
  carlo.
\newblock \emph{Available at SSRN 2328709}, 2013.

\bibitem[Breeden and Litzenberger(1978)]{breeden}
D.~T. Breeden and R.~H. Litzenberger.
\newblock Prices of state-contingent claims implicit in option prices.
\newblock \emph{Journal of business}, pages 621--651, 1978.

\bibitem[Broadie and Cao(2008)]{Cao}
M.~Broadie and M.~Cao.
\newblock Improved lower and upper bound algorithms for pricing american
  options by simulation.
\newblock \emph{Quantitative Finance}, 8\penalty0 (8):\penalty0 845--861, 2008.

\bibitem[Buehler et~al.(2019)Buehler, Gonon, Teichmann, and Wood]{buehler}
H.~Buehler, L.~Gonon, J.~Teichmann, and B.~Wood.
\newblock Deep hedging.
\newblock \emph{Quantitative Finance}, pages 1--21, 2019.

\bibitem[Carr and Wu(2013)]{carr}
P.~Carr and L.~Wu.
\newblock Static hedging of standard options.
\newblock \emph{Journal of Financial Econometrics}, 12\penalty0 (1):\penalty0
  3--46, 2013.

\bibitem[Carriere(1996)]{carriere1996valuation}
J.~F. Carriere.
\newblock Valuation of the early-exercise price for options using simulations
  and nonparametric regression.
\newblock \emph{Insurance: mathematics and Economics}, 19\penalty0
  (1):\penalty0 19--30, 1996.

\bibitem[Derman et~al.()Derman, Ergener, and Kani]{derman}
E.~Derman, D.~Ergener, and I.~Kani.
\newblock Static options replication.
\newblock \emph{Journal of Derivatives}, 2\penalty0 (4).

\bibitem[Fang and Oosterlee(2008)]{COS}
F.~Fang and C.~W. Oosterlee.
\newblock A novel pricing method for european options based on fourier-cosine
  series expansions.
\newblock \emph{SIAM Journal on Scientific Computing}, 31\penalty0
  (2):\penalty0 826--848, 2008.

\bibitem[Feng et~al.(2016)Feng, Jain, Karlsson, Kandhai, and Oosterlee]{qian}
Q.~Feng, S.~Jain, P.~Karlsson, D.~Kandhai, and C.~W. Oosterlee.
\newblock Efficient computation of exposure profiles on real-world and
  risk-neutral scenarios for bermudan swaptions.
\newblock \emph{Journal of Computational Finance}, 20\penalty0 (1):\penalty0
  139--172, 2016.

\bibitem[Glasserman and Yu(2004)]{glasserman2004simulation}
P.~Glasserman and B.~Yu.
\newblock Simulation for american options: Regression now or regression later?
\newblock In \emph{Monte Carlo and Quasi-Monte Carlo Methods 2002}, pages
  213--226. Springer, 2004.

\bibitem[Halperin(2019)]{halperin}
I.~Halperin.
\newblock The qlbs q-learner goes nuqlear: fitted q iteration, inverse rl, and
  option portfolios.
\newblock \emph{Quantitative Finance}, 19\penalty0 (9):\penalty0 1543--1553,
  2019.

\bibitem[Haugh and Kogan(2004)]{haugh}
M.~B. Haugh and L.~Kogan.
\newblock Pricing american options: a duality approach.
\newblock \emph{Operations Research}, 52\penalty0 (2):\penalty0 258--270, 2004.

\bibitem[Hornik et~al.(1989)Hornik, Stinchcombe, and White]{hornick}
K.~Hornik, M.~Stinchcombe, and H.~White.
\newblock Multilayer feedforward networks are universal approximators.
\newblock \emph{Neural networks}, 2\penalty0 (5):\penalty0 359--366, 1989.

\bibitem[Jain and Oosterlee(2015)]{sgbm}
S.~Jain and C.~W. Oosterlee.
\newblock The stochastic grid bundling method: Efficient pricing of bermudan
  options and their greeks.
\newblock \emph{Applied Mathematics and Computation}, 269:\penalty0 412--431,
  2015.

\bibitem[Jain et~al.(2019{\natexlab{a}})Jain, Karlsson, and Kandhai]{kva}
S.~Jain, P.~Karlsson, and D.~Kandhai.
\newblock Kva, mind your p's and q's!
\newblock \emph{Wilmott}, 2019\penalty0 (102):\penalty0 60--73,
  2019{\natexlab{a}}.

\bibitem[Jain et~al.(2019{\natexlab{b}})Jain, Leitao, and
  Oosterlee]{rollingAdjoints}
S.~Jain, {\'A}.~Leitao, and C.~W. Oosterlee.
\newblock Rolling adjoints: fast greeks along monte carlo scenarios for
  early-exercise options.
\newblock \emph{Journal of Computational Science}, 33:\penalty0 95--112,
  2019{\natexlab{b}}.

\bibitem[Kingma and Ba(2014)]{adam}
D.~P. Kingma and J.~Ba.
\newblock Adam: A method for stochastic optimization.
\newblock \emph{arXiv preprint arXiv:1412.6980}, 2014.

\bibitem[Kohler et~al.(2010)Kohler, Krzy{\.z}ak, and Todorovic]{kohler}
M.~Kohler, A.~Krzy{\.z}ak, and N.~Todorovic.
\newblock Pricing of high-dimensional american options by neural networks.
\newblock \emph{Mathematical Finance: An International Journal of Mathematics,
  Statistics and Financial Economics}, 20\penalty0 (3):\penalty0 383--410,
  2010.

\bibitem[Lapeyre and Lelong(2019)]{lapeyre2019neural}
B.~Lapeyre and J.~Lelong.
\newblock Neural network regression for bermudan option pricing.
\newblock \emph{arXiv preprint arXiv:1907.06474}, 2019.

\bibitem[Leshno et~al.(1993)Leshno, Lin, Pinkus, and Schocken]{leshno}
M.~Leshno, V.~Y. Lin, A.~Pinkus, and S.~Schocken.
\newblock Multilayer feedforward networks with a nonpolynomial activation
  function can approximate any function.
\newblock \emph{Neural networks}, 6\penalty0 (6):\penalty0 861--867, 1993.

\bibitem[Liu et~al.(2019{\natexlab{a}})Liu, Borovykh, Grzelak, and
  Oosterlee]{liu2}
S.~Liu, A.~Borovykh, L.~A. Grzelak, and C.~W. Oosterlee.
\newblock A neural network-based framework for financial model calibration.
\newblock \emph{arXiv preprint arXiv:1904.10523}, 2019{\natexlab{a}}.

\bibitem[Liu et~al.(2019{\natexlab{b}})Liu, Oosterlee, and Bohte]{liu}
S.~Liu, C.~W. Oosterlee, and S.~M. Bohte.
\newblock Pricing options and computing implied volatilities using neural
  networks.
\newblock \emph{Risks}, 7\penalty0 (1):\penalty0 16, 2019{\natexlab{b}}.

\bibitem[Longstaff and Schwartz(2001)]{longstaff2001valuing}
F.~A. Longstaff and E.~S. Schwartz.
\newblock Valuing american options by simulation: a simple least-squares
  approach.
\newblock \emph{The review of financial studies}, 14\penalty0 (1):\penalty0
  113--147, 2001.

\bibitem[Pellizzari(2005)]{pellizzari}
P.~Pellizzari.
\newblock Static hedging of multivariate derivatives by simulation.
\newblock \emph{European Journal of Operational Research}, 166\penalty0
  (2):\penalty0 507--519, 2005.

\bibitem[Pelsser(2003)]{pelsser2}
A.~Pelsser.
\newblock Pricing and hedging guaranteed annuity options via static option
  replication.
\newblock \emph{Insurance: Mathematics and Economics}, 33\penalty0
  (2):\penalty0 283--296, 2003.

\bibitem[Pelsser and Schweizer(2016)]{pelsser}
A.~Pelsser and J.~Schweizer.
\newblock The difference between lsmc and replicating portfolio in insurance
  liability modeling.
\newblock \emph{European actuarial journal}, 6\penalty0 (2):\penalty0 441--494,
  2016.

\bibitem[Reisinger and Wittum(2007)]{reisinger}
C.~Reisinger and G.~Wittum.
\newblock Efficient hierarchical approximation of high-dimensional option
  pricing problems.
\newblock \emph{SIAM Journal on Scientific Computing}, 29\penalty0
  (1):\penalty0 440--458, 2007.

\bibitem[Rogers(2002)]{rogers}
L.~C. Rogers.
\newblock Monte carlo valuation of american options.
\newblock \emph{Mathematical Finance}, 12\penalty0 (3):\penalty0 271--286,
  2002.

\bibitem[Tsitsiklis and Van~Roy(2001)]{tsitsiklis2001regression}
J.~N. Tsitsiklis and B.~Van~Roy.
\newblock Regression methods for pricing complex american-style options.
\newblock \emph{IEEE Transactions on Neural Networks}, 12\penalty0
  (4):\penalty0 694--703, 2001.

\end{thebibliography}
\appendix
\section{}\label{Ax}
When the underlying asset prices follow a GBM process, we want to evaluate the following expression,
$$ \E\left[\varphi \left(\mathbf{w}^{\top}_{i}\log\left(\BS_{t_m}\right) + b_{i} \right)\mid\BS_{t_{m-1}}\right].$$

We begin by noting that conditioned on $\F_{t_{m-1}},$ the random vector $\log\left(\BS_{t_m}\right)$ ---under the risk-neutral measure--- has a multivariate normal distribution with the mean vector  given by 

$$
\boldsymbol{\mu}:=\begin{pmatrix}
\mu_1\\
\vdots\\
\mu_d
\end{pmatrix}:= \log\left(\BS_{t_{m-1}}\right) + \begin{pmatrix}
r-\frac{1}{2}\sigma_1^2\\
\vdots\\
r-\frac{1}{2}\sigma_d^2
\end{pmatrix} \Delta t,
$$

and the covariance matrix by,

$$
\Sigma := \begin{pmatrix}
\sigma_1^2&\rho_{12}\sigma_1\sigma_2&\hdots&\rho_{1d}\sigma_1\sigma_d\\
\rho_{21}\sigma_2\sigma_1& \sigma_2^2 &\hdots&\rho_{2d}\sigma_2\sigma_d\\
\vdots&\vdots&\ddots&\vdots\\
\rho_{d1}\sigma_d\sigma_1&\rho_{d2}\sigma_d\sigma_2&\hdots&\sigma_d^2
\end{pmatrix} \Delta t,
$$

where $\Delta t = t_{m}-t_{m-1}.$

We denote the conditional weighted sum of the log asset prices  by a random variable $Y.$ Then $Y$ has a normal distribution with mean,

$$
\mu_Y := \mathbf{w}_i^{\top} \boldsymbol{\mu} ,
$$

and variance,

$$
\sigma_Y :=  \mathbf{w}_i^{\top}  \Sigma \mathbf{w}_i
$$

We need to then determine

\begin{eqnarray}\nonumber
 \E\left[\varphi \left(\mathbf{w}^{\top}_{i}\log\left(\BS_{t_m}\right) + b_{i} \right)\mid\BS_{t_{m-1}}\right] &=& \E\left[\max\left(Y + b_{i},0 \right)\right]\\\nonumber
 &=&\E\left[\max\left(\tilde{Y},0 \right)\right],
\end{eqnarray}

where $\tilde{Y}$ is a normal with mean $\mu_{\tilde{Y}}=\mu_Y+b_i,$ and variance $\sigma_Y.$The expression is straightforward to evaluated and is given by

\begin{eqnarray}
\E\left[\max\left(\tilde{Y},0 \right)\right]&=&\frac{1}{\sqrt{2\pi\sigma_Y^2}}\int_{0}^{\infty} y e^{\frac{-\left(y-\mu_{\tilde{Y}}\right)^2}{2\sigma_Y^2}}\ dy\\
&=&\frac{1}{\sqrt{2\pi\sigma_Y^2}}\int_{-\mu_{\tilde{Y}}}^{\infty} \left(x+\mu_{\tilde{Y}}\right) e^{\frac{-x^2}{2\sigma_Y^2}} \ dx\\
&=&\frac{1}{\sqrt{2\pi\sigma_Y^2}}\int_{-\mu_{\tilde{Y}}}^{\infty} \left(x\right) e^{\frac{-x^2}{2\sigma_Y^2}}\ dx +\frac{1}{\sqrt{2\pi\sigma_Y^2}} \int_{-\mu_{\tilde{Y}}}^{\infty} \left(\mu_{\tilde{Y}}\right) e^{\frac{-x^2}{2\sigma_Y^2}}\ dx\\
&=& \frac{\sigma_Y}{\sqrt{2 \pi}} e^{\frac{-\mu^2_{\tilde{Y}}}{2\sigma_Y^2}} + \mu_{\tilde{Y}}\left(1-F_{\tilde{Y}}(-\mu_{\tilde{Y}})\right),
\end{eqnarray}

where $F_{\tilde{Y}}$ is the cdf of $\tilde{Y}.$

\end{document}